\documentclass[a4paper,preprintnumbers,11pt]{article}
\pdfoutput=1 

\usepackage{jheppub} 

\usepackage[T1]{fontenc} 

\usepackage{float}
\usepackage{mathtools}
\usepackage{amsfonts} 
\usepackage{amssymb} 
\usepackage{amsmath} 
\usepackage{graphicx} 
\usepackage{subfigure} 
\usepackage{array} 
\usepackage{dcolumn} 
\usepackage{bm} 
\usepackage{latexsym} 
\usepackage{longtable} 
\usepackage{hyperref} 
\usepackage{verbatim}
\usepackage{epsfig}
\usepackage{slashed}
\usepackage{color}
\newcommand{\wt}[0]{\widetilde}

\newcommand{\cK}[0]{\mathcal K}

\newcommand{\cM}[0]{\mathcal M}
\newcommand{\cO}[0]{\mathcal O}

\newcommand{\df}[0]{\mathrm{df}}
\newcommand{\K}[0]{\mathcal K}

\newcommand{\iso}[0]{{\rm iso}}

\newcommand{\Kiso}[0]{{\cK_{\df,3}^{\iso}}}
\newcommand{\Kdfone}[0]{{\cK_{\df,3}^{\iso,1}}}
\newcommand{\Kdf}[0]{{\cK_{\df,3}}}

\newcommand{\PV}[0]{{\mathrm{PV}}}

\newcommand{\Kdfzero}[0]{{\cK_{\df,3}^{\iso,0}}}

\newcommand{\HSBS}[0]{Hansen:2016ync}

\newcommand{\BHSQC}[0]{Briceno:2017tce}
\newcommand{\BHSnum}[0]{Briceno:2018mlh}
\newcommand{\BHSK}[0]{Briceno:2018aml}
\newcommand{\HSQCa}[0]{Hansen:2014eka}
\newcommand{\HSQCb}[0]{Hansen:2015zga}
\newcommand{\dwave}[0]{Blanton:2019igq}
\newcommand{\AkakiBS}[0]{Meissner:2014dea}
\newcommand{\Akakia}[0]{Hammer:2017uqm}
\newcommand{\Akakib}[0]{Hammer:2017kms}

\newcommand{\MD}[0]{Mai:2017bge}
\newcommand{\MDpi}[0]{Mai:2018djl}
\newcommand{\Akakinum}[0]{Doring:2018xxx}

\newcommand{\HSrev}[0]{Hansen:2019nir}
\newcommand{\BRSd}[0]{Blanton:2019igq}

\newcommand{\KSS}[0]{Kim:2005gf}

\newcommand{\PhillipsOrig}[0]{Phillips:1968zze}
\newcommand{\PhillipsBHK}[0]{Bedaque:1999ve}
\newcommand{\BHSSunit}[0]{Briceno:2019muc}

\title{ Numerical exploration of three relativistic particles in a finite volume including two-particle resonances and bound states }

\author[1]{Fernando Romero-L\'opez}
\affiliation[1]{IFIC, CSIC-Universitat de Val\`encia, 46980 Paterna, Spain}

\author[2]{, Stephen R. Sharpe}
\affiliation[2]{Physics Department, University of Washington, Seattle, WA 98195-1560, USA}

\author[2]{, Tyler D. Blanton}

\author[3,4]{, Ra\'ul A. Brice\~no}
\affiliation[3]{Thomas Jefferson National Accelerator Facility, 12000 Jefferson Avenue, Newport News, Virginia 23606, USA}
\affiliation[4]{ Department of Physics, Old Dominion University, Norfolk, Virginia 23529, USA}

\author[5]{ and Maxwell T. Hansen}
\affiliation[5]{Theoretical Physics Department, CERN, 1211 Geneva 23, Switzerland}

\emailAdd{fernando.romero@uv.es}
\emailAdd{srsharpe@uw.edu}
\emailAdd{blanton1@uw.edu}
\emailAdd{rbriceno@jlab.org}
\emailAdd{maxwell.hansen@cern.ch}

\abstract{In this work, we use an extension of the quantization condition, given in Ref.~\cite{\HSQCa}, to numerically explore the finite-volume spectrum of three relativistic particles, in the case that two-particle subsets are either resonant or bound.
The original form of the relativistic three-particle quantization condition was derived under 
a technical assumption on the two-particle K matrix that
required the absence of two-particle bound states or narrow two-particle resonances.
Here we describe how this restriction can be lifted in a simple way
using the freedom in the definition of the K-matrix-like quantity that enters the quantization condition.
With this in hand, we extend previous numerical studies of the quantization condition
to explore the finite-volume signature for a variety of two- and three-particle interactions.
We determine the spectrum
for parameters such that the system contains both dimers (two-particle bound states) 
and one or more trimers (in which all three particles are bound),
and also for cases where the two-particle subchannel is resonant.
 We also show how the quantization condition provides a tool for 
determining infinite-volume dimer-particle scattering amplitudes for energies below the dimer breakup.
 We illustrate this for a series of examples, including one that parallels physical deuteron-nucleon scattering.
All calculations presented here are restricted to the case of three identical scalar particles.
}

\begin{document} 
\preprint{\begin{tabular}{r}CERN-TH-2019-129\\JLAB-THY-19-3011\end{tabular}}

\maketitle
\flushbottom

\section{Introduction \label{sec:intro}}

Lattice quantum chromodynamics (LQCD), in which QCD correlators are estimated numerically via Monte Carlo importance sampling of the path integral, has proven to be a powerful tool for determining low-energy 
properties of hadrons. 
Currently, one of the major frontiers of numerical LQCD is the calculation of few-hadron observables.
In particular, there has been  substantial recent
 progress in the determination of scattering amplitudes, 
including cases for which multiple channels are open and 
couple to underlying resonances~\cite{Wilson:2015dqa, Briceno:2016mjc, Brett:2018jqw, Andersen:2017una, Guo:2018zss, Andersen:2018mau, Dudek:2014qha, Dudek:2016cru, Woss:2018irj, Woss:2019hse,Helmes:2018nug,Liu:2016cba,Helmes:2017smr,Helmes:2015gla,Werner:2019hxc,Culver:2019qtx}.\footnote{%
See Ref.~\cite{Briceno:2017max} for a recent review.}
 These studies rely on formalism that maps   quantities obtained via LQCD,
namely finite-volume observables, to infinite-volume amplitudes~\cite{Luscher:1986n2,Luscher:1991n1,Kari:1995, Kim:2005gf, He:2005ey, Bernard:2010, Hansen:2012tf, Briceno:2012yi, Briceno:2014oea, Romero-Lopez:2018zyy}.  
 
 Presently, one of the primary limitations on the study of resonances and light nuclei is the absence 
 of a complete formalism that can provide such  a mapping for energies above  three-particle 
 production thresholds. 
 In fact, finite-volume spectra are already being obtained above 
 three-particle  thresholds using  a large basis of interpolators including 
those built from three single-hadron operators, each projected to definite momentum~\cite{Cheung:2017tnt, Woss:2019hse,Horz:2019rrn}.\footnote{%
Similar work has also been done in the $\varphi^4$ theory~\cite{Romero-Lopez:2018rcb}.} 
Without a three-particle formalism, the information contained in these spectra is inaccessible. This was recently emphasized in a USQCD whitepaper: ``\emph{...the development of a rigorous three-body (and higher) formalism is vital to have confidence in the calculations of high-lying resonances}''~\cite{Detmold:2019ghl}.

While a complete formalism is not yet in place,
there has been considerable progress in this direction, following three approaches.
The first is based on a generic
relativistic effective field theory (the relativistic field theory or RFT approach)~\cite{\HSQCa,\HSQCb,\BHSQC,\BHSnum,\BHSK,\dwave}, 
the second uses non-relativistic effective field theory (the NREFT method)~\cite{\Akakia,\Akakib,\Akakinum},
and the third applies unitary constraints in finite volumes 
(the FVU or finite-volume unitarity approach)~\cite{\MD,\MDpi}.\footnote{%
It is worth emphasizing that efforts to constrain infinite-volume three-particle amplitudes from 
finite-volume LQCD results has partially motivated several 
infinite-volume studies~\cite{Mai:2017vot, Mikhasenko:2019vhk, Jackura:2018xnx, Briceno:2019muc, Jackura:2019bmu}.}
For a recent review, including a discussion of the relation between the different formalisms,
see Ref.~\cite{\HSrev}.  At present, the only formalism that is both fully relativistic and 
incorporates partial-wave mixing (both due to the physical three-body dynamics and
 the reduction of rotational symmetry in a finite volume) is the RFT approach. 
 In this work we focus on this approach and extend its range of applicability.   
 
In the original derivation, given in Ref.~\cite{\HSQCa}, it was necessary to assume that
a quantity closely related to the two-particle K matrix
had no singularities in the kinematic region of interest. This implies that the formalism cannot be used if the two-particle subchannels contain
resonances that are narrow enough to induce such singularities, 
or bound states, which generically give poles in the K-matrix-like quantity.
The formalism also breaks down if
the K matrix contains poles above threshold that do not correspond to a physical phenomenon,
as occurs, for example, when the corresponding phase shift passes through $\pi/2$ from above.
These restrictions are a major practical shortcoming of the original RFT approach.
They are also surprising, as the problematic poles do not, in general, correspond to physical quantities.
All such technical restrictions were lifted by a recent extension of the formalism given in Ref.~\cite{\BHSK},
but at the cost of including an unphysical channel at intermediate stages for each bound state or resonance,
making the approach cumbersome in practical implementations.

In prior studies, we have explored the numerical implementation of the original formalism
of Ref.~\cite{\HSQCa} in simple limits. 
First, in Ref.~\cite{\BHSnum} we considered the low-energy, isotropic approximation,
in which scattering in two-particle subchannels is dominated by the $s$-wave, and the
three-particle scattering quantity, $\Kdf$, is independent of the spectator momentum. 
Second, in Ref.~\cite{\dwave}, we considered the case in which $d$-wave scattering, 
the dominant subleading partial wave for identical particle systems at low energies, 
was no longer negligible, with corresponding nonisotropic contributions added to $\Kdf$.
The latter investigation demonstrated that higher partial wave systems 
can indeed be implemented numerically using the RFT approach. 
However, because of the above-mentioned restrictions, in both studies we
were only able to consider dynamics where two-particle subsystems did not have bound states or resonances. Three-particle bound states, for which no issues arise, are considered in Refs.~\cite{\HSBS,\BHSnum,\dwave}.

In this work we describe and implement  an alternative modification of the formalism of Ref.~\cite{\HSQCa} that removes the restriction on K matrix poles.
This new approach does not require the introduction of unphysical channels,
and is thus much simpler than that given in Ref.~\cite{\BHSK}.
Indeed, the numerical implementation of the new approach requires only slight modifications compared to that for the original formalism,
so the methods of Refs.~\cite{\BHSnum, \dwave} can be used with minimal change.
We stress that, 
by allowing for a general two-particle K matrix, our improvement of the original formalism brings 
it to the same status in this regard as the NREFT and FVU approaches~\cite{\Akakib,\MD,\MDpi,\Akakinum}, since the latter do not require restrictions on the two-particle K matrix.
We also note that, since the RFT formalism is valid for arbitrary partial waves, and has been
implemented for combined $s$- and $d$-waves, we are able to consider three-particle systems
not previously addressed in the literature. Finally we comment that this new method is also of relevance for $2 \to 3$ scattering, already considered in the context of the RFT formalism in Ref.~\cite{\BHSQC}. Further work is required to relate the present ideas to the coupled-channel formalism of Ref.~\cite{\BHSQC}.

The main purpose of the present note is to show examples of the results that are obtained
using the modified formalism. In a companion paper~\cite{inprog} we will describe in detail why the
modified formalism is valid, as well as the relation to the more complicated approach of Ref.~\cite{\BHSK}.

The finite- to infinite-volume relation consists of two steps. The first 
uses a three-particle quantization condition to relate the finite-volume spectrum, $E_n(L)$, to an intermediate,
scheme-dependent, infinite-volume three-particle scattering quantity, $\Kdf$,
while the second requires solving infinite-volume integral equations to relate $\Kdf$ to the physical scattering amplitude, $\cM_3$.
In Ref.~\cite{\BHSnum}, we implemented both steps, although the latter only below and at the
three-particle threshold. 
Here, in the interests of brevity and clarity,  we mainly consider the first step, $E_n(L) \leftrightarrow \Kdf$, and only indirectly explore some consequences of the second: $\Kdf \leftrightarrow \mathcal M_3$.

As in all work to date, we restrict ourselves  to the case of identical
scalar particles. Thus, strictly speaking, the formalism applies in QCD
only to three identical pions (e.g.~$\pi^+ \pi^+ \pi^+$).
We further restrict ourselves to theories with a G-parity-like $\mathbb Z_2$ symmetry that conserves
particle-number modulo two (as is the case for three pions in the isospin-symmetric limit).\footnote{%
The presence of a $\mathbb Z_2$ symmetry was assumed in the original derivation~\cite{\HSQCa,\HSQCb},
and has also been assumed in all numerical investigations so far. The formalism for a theory without
a $\mathbb Z_2$ symmetry has been developed~\cite{\BHSQC}, but has not yet been numerically implemented.
}
This forbids $2\to 3$ transitions, which, in the energy range we consider,
leads effectively to particle-number conservation. Furthermore, since many features of the 
finite-volume spectrum are determined by kinematics, we expect that the examples we show will shed
light on the three-nucleon system.
To illustrate this, in Sec.~\ref{sec:nD} we choose parameters to mimic the $n n p$  or $n p p$ three-nucleon systems,
with a two-particle bound state (called a dimer state) having the same binding energy as the deuteron, and
the three-particle bound state (referred to as a trimer state) having the same binding energy as the triton or helium-3.
This enables us to study a toy version of nucleon-deuteron scattering and reproduce the well known Phillips line of nuclear effective field theory \cite{\PhillipsOrig, \PhillipsBHK}.

{
This example brings out an important general point.
The three-particle quantization condition was developed in order to map finite-volume energies into infinite-volume scattering observables. 
It turns out, however, that
the formalism also predicts the properties of two- and three-particle resonances and bound states, given a description of the microscopic physics as encoded in the two-particle K matrix and $\Kdf$. This aspect is independent of the finite-volume and emerges because the integral equations relating $\Kdf$ to $\mathcal M_3$ exactly solve the unitarity constraints on the scattering amplitude \cite{\BHSSunit}. 
This alternative application of the formalism was already used
in Ref.~\cite{\BHSnum}, where we extracted the vertex function of a trimer
 and found good agreement with the Efimov wave-function.  Similarly, in Ref.~\cite{\dwave}, we used the approach to give evidence for a trimer that is bound primarily by attractive
$d$-wave interactions.}

The remainder of this paper is organized as follows. We begin, in Sec.~\ref{sec:recap}, by providing a brief recap of the essential components of the formalism of Ref.~\cite{\HSQCa}. This is followed in Sec.~\ref{sec:generalizing} by a description of the modified formalism that allows for the study of three-body states with either resonant or bound subsystems. In Sec.~\ref{sec:numerical} we illustrate the power of this formalism by applying it to various examples: First, in Sec.~\ref{sec:largea0} we determine the finite-volume spectrum for systems with dimers; then, in Sec.~\ref{sec:dimerpart}, we evaluate the finite-volume spectrum below the three-particle threshold at large volumes, in order to determine the particle-dimer
scattering amplitude for a range of two-body scattering lengths;
third, in Sec.~\ref{sec:nD}, we tune $\Kiso$ and the two-body parameters to determine the neutron-deuterium scattering amplitude in a toy model without spin or isospin; next in Sec.~\ref{sec:BWres} we present the three-particle spectrum in the case of a two-particle resonance; 
and, finally, in Sec.~\ref{sec:numdwave}, we consider the implication of including $d$-wave dimers. 
We present concluding remarks in Sec.~\ref{sec:conclusions}.
We also include two appendices. In Appendix~\ref{app:IPV} we explore the role that the scheme-dependence of $\Kdf$ plays in determining the finite-volume energy spectrum.
In Appendix~\ref{app:NREFT} we explain how the predictions for the particle-dimer scattering length 
are obtained in the NREFT framework.

\section{Recap of the quantization condition and its approximations \label{sec:recap}}

In the presence of a $\mathbb Z_2$ symmetry, 
the finite-volume spectrum is determined by the solutions, $E=E_n(L)$ with $n=0,1,2,\ldots$, to the quantization
condition~\cite{\HSQCa}\footnote{%
The quantization condition holds up to exponentially-suppressed corrections, 
scaling as $e^{-mL}$, which are  assumed negligible, and ignored, throughout this work.}
\begin{equation}
\det\left[ F_3(E,L)^{-1}+ \Kdf(E) \right] = 0
\,.
\label{eq:QC}
\end{equation}
Here $L$ is the linear extent of the finite cubic spatial volume, effected by applying periodic boundary conditions to the fields defining the theory.
Although the formalism holds for arbitrary total three-momentum, $\vec P$,
we consider here only the case in which $\vec P=0$,
so that the total energy $E$ is also the center-of-mass energy for the three particles. 
The quantization condition is valid for $E<5m$, i.e. for energies below the five-particle production threshold.

The second term appearing in the determinant, $\Kdf$, is the aforementioned, scheme-dependent, infinite-volume scattering quantity.
It is a smooth, real function of the kinematic variables describing
three-to-three scattering and can be understood as the short-distance piece of the three-body interaction.\footnote{%
The label ``df'' denotes ``divergence-free'', and indicates that kinematical divergences, present in the three-particle amplitude $\cM_3$, are removed in the definition of $\Kdf$.
For more details, see Refs.~\cite{\HSQCa,\HSQCb}. 
 We note that, just as for the two-body K matrix, $\Kdf$ can have poles induced
 by the dynamics, although we do not consider this possibility here.}
The first term in the determinant, $F_3$, depends on the physical two-particle  scattering amplitude, $\mathcal M_2$ (or equivalently, through a straightforward algebraic relation, on the two-particle K matrix, $\mathcal K_2$) and on known geometric 
functions that depend on the box shape and size.

In Eq.~(\ref{eq:QC}),  both $F_3$ and $\Kdf$ are written as
infinite-dimensional matrices acting on the space of three, on-shell particles
in finite volume. Each object carries two copies of the index set $\vec k,\ell, m$, where $\vec k=2\pi \vec n/L$ is a finite-volume momentum, given in terms of a 3-vector of integers, $\vec n$,
while $\ell$ and $m$ are angular-momentum indices. 
The set-up is that $\vec k$ is the momentum of one of the three on-shell particles, referred to as the
spectator, while $\ell$ and $m$ describe the angular momentum of the other two in their 
center-of-mass (c.m.) frame. 
$F_3$ is intrinsically a finite-volume quantity, and thus comes always in matrix form.
By contrast, $\Kdf$ is an infinite-volume quantity depending on continuous momentum coordinates and the matrix version is defined by sampling the function at a discrete set of kinematics, dictated by the volume.%
\footnote{%
Explicit examples of how to carry out this restriction are given in Ref.~\cite{\dwave}.}

As we describe below, the spectator-momentum index, $\vec k$, is cut off by a function
$H(\vec k)$ that serves as an ultraviolet regulator. Thus the matrices in Eq.~(\ref{eq:QC}) are
infinite only due to their angular-momentum indices. One can expand 
the two- and three-particle K matrices $\K_2$ and $\Kdf$ about threshold, 
as explained in Refs.~\cite{\BHSnum,\dwave}. This leads to a systematic truncation scheme
in which $\ell \le \ell_{\rm max}$ in all quantities entering the quantization condition,
including the kinematic functions~\cite{\HSQCa}. 
Since both spectator momentum and angular-momentum index sums are truncated,
the problem reduces to one involving finite matrices,
suitable for numerical implementation.
In this work we will present results for both $\ell_{\rm max}=0$ ($s$-wave only) and $\ell_{\rm max} = 2 $ ($s$- and $d$-wave mixing).\footnote{%
Odd $\ell$ give vanishing contibutions due to the exchange symmetry of the identical scalar particles.}
%

\subsection{The $s$-wave-only approximation: $\ell_{\rm max}=0$ \label{sec:swave}}

The general expression for $F_3$ is given in Ref.~\cite{\HSQCa}. 
Here we recall the form only for the simplest case, $\ell_{\rm max}=0$,
which is the choice we use in most of the numerical explorations described below.
In this limit, the index space reduces from $\vec k, \ell, m$, to the discretized momentum, $\vec k$. Denoting the $s$-wave-only version of $F_3$ by $F_3^s$, we recall that the latter is given by
\begin{equation}
L^3 F_3^s \equiv \frac{\wt F^s}{3} - \wt F^s \frac1{1/\wt \K_2^s + \wt F^s + \wt G^s} \, \wt F^s
\,,
\label{eq:F3s}
\end{equation}
where $\wt F^s$ and $\wt G^s$ are geometric matrices in the space of the spectator momentum,
\begin{align}
[\wt F^s]_{kp} & \equiv \frac{\delta_{kp}}{2} \frac{H(\vec k)}{2\omega_k}
\bigg [ \frac1{L^3}\sum_{\vec a} - \PV\int \frac{d^3 a}{(2\pi)^3} \bigg ]
\frac{H_2(\vec a, \vec b)}{2 \omega_a   2\omega_b (E-\omega_k-\omega_a-\omega_b)}\,,
\label{eq:Fts}
\\
[\wt G^s]_{kp} & \equiv \frac{H(\vec k) H(\vec p\,)}{ L^3 2  \omega_k 2\omega_p ( b^2 - m^2)}.
\label{eq:Gts}
\end{align}
The sum in Eq.~(\ref{eq:Fts}) runs over all finite-volume momenta, i.e.~over all  $\vec a=(2\pi/L) \vec n_a$ where $\vec n_a$ is a 3-vector of integers. As we set $\vec P=0$, the third particle carries momentum $\vec b \equiv -\vec a - \vec k$. On-shell energies are denoted $\omega$, for example
$\omega_k \equiv \sqrt{m^2 + \vec k^2}$, with $m$ the particle mass. 
The explicit form of the cutoff function $H(\vec k)$ is given in Refs.~\cite{\HSQCa,\BHSQC,\BHSnum} and
is not repeated here, except to note that we always take $\alpha=-1$
for the parameter in the cutoff function. [See Eq.~(A3) of Ref.~\cite{\BHSQC}.]

The sum-integral difference in Eq.~(\ref{eq:Fts}) is regulated in the ultraviolet
by the function $H_2$, for which there is considerable freedom.
In this work 
we use the ``KSS'' form~\cite{\KSS}, 
$H_2(\vec a, \vec b) = \exp[- \alpha_{\text{KSS}} (a^{* 2} - q^{* 2}) ]$, 
explained in detail in Appendix B of Ref.~\cite{\BHSnum}.

We note that while $\wt F^s$ is the same as in Ref.~\cite{\BHSnum}, 
$\wt G^s$ differs---here we use its relativistic form,
since this leads to a Lorentz invariant $\Kdf$. This invariance plays a role when expanding this function about three-particle threshold. 
Unlike for $\wt G^s$, it is not necessary that the denominator
in $\wt F^s$ be relativistically invariant. This is because replacing
$2\omega_b(E-\omega_k-\omega_a-\omega_b)$ with $(b^2-m^2)$ leads only to
an exponentially suppressed change to $\wt F^s$.

The final ingredient needed for $\wt F_3^s$ is the two-particle $s$-wave K matrix, or, equivalently,
the $s$-wave phase shift $\delta_s$. This appears in the  diagonal matrix
\begin{align}
[1/{\wt {\K}_2^s}]_{kp} & \equiv  \delta_{kp} \big (1/{\wt \K_2^s(\vec k)} \big)\,,
\label{eq:K2tinv}
\\
{\wt \K_2^s(\vec k)} & \equiv
\frac{32 \pi \omega_k E_{2,k}^*}{
q_{2,k}^* \cot\delta_s(q_{2,k}^*) + |q_{2,k}^*| [1-H(\vec k)]}\,,
\label{eq:K2t}
\end{align}
where
\begin{equation}
E_{2,k}^{*2} = (E-\omega_k)^2-\vec k^2 \ \ {\rm and}\ \
q_{2,k}^{*2}= \frac{E_{2,k}^{*2}}4-m^2
\label{eq:E2k}
\end{equation}
are the total squared energy and particle momentum
in the c.m.~frame of the nonspectator pair,
and $\delta_s$ the $s$-wave phase shift.
Were it not for the second term in the denominator of Eq.~(\ref{eq:K2t}), $\wt \K_2^s$ would
simply equal $\K_2/(2\omega_k)$, where
\begin{equation}
 \K_2^s(\vec k)   \equiv
\frac{16 \pi   E_{2,k}^*}{
q_{2,k}^* \cot\delta_s(q_{2,k}^*)   }\,,
\end{equation}
is one standard choice for the definition of the K matrix. Indeed, the equivalence does hold above threshold (i.e.~for $E_{2,k}^* > 2m $), where $H(\vec k)=1$.
The second term is essential, however, for the derivation of
Ref.~\cite{\HSQCa}, and implies that $\wt \K_2^s$ is scheme-dependent below threshold. 

In what follows, we make use of two parametrizations of the phase shift. 
The first is a low-energy expansion, commonly referred to as the effective range expansion
(ERE),
which for the $s$-wave can be written as
\begin{align}
q_{2,k}^* \cot \delta_s(q_{2,k}^*) = - \frac{1}{a_0} + \frac12 r_0 q_{2,k}^{*2}  + \mathcal O(q_{2,k}^{*4}) \,,
\label{eq:effrange}
\end{align}
where $a_0$ and $r_0$ are the scattering length and the effective range, respectively. 
In numerical explorations considered below, we will only consider examples with $r_0=0$. 
Our second choice is the Breit-Wigner form, commonly used when
a narrow resonance couples to a two-body system. 
For an $s$-wave resonance, this  can be written 
\begin{equation}
\tan \delta_{\rm BW}(q_{2,k}^*) = \frac{E_{2,k}^{*} \, \Gamma(E_{2,k}^{*})}{m_R^2 - E_{2,k}^{*2}} 
\ \ \ \ \   \mathrm{with} \ \ \ \ \   \Gamma(E_{2,k}^{*}) = \frac{g^2}{6\pi} \frac{m_R^2}{E_{2,k}^{*2} } q_{2,k}^*\,,
\label{eq:BW}
\end{equation} 
where $m_R$ is the resonance mass and $g$ its coupling to the two-particle channel.
We describe below one example of the finite-volume three-particle
spectrum in the presence of such a resonant two-body interaction.
 We note that the Breit-Wigner form is similar to the truncated effective range
expansion,
\begin{equation}
q_{2,k}^* \cot \delta_{\rm BW}(q_{2,k}^*) = \frac{m_R^2 - E_{2,k}^{*2}} {E_{2,k}^{*}  }  \frac{6\pi}{g^2} \frac{E_{2,k}^{*2} }{m_R^2} = E_{2,k}^{*}  [A + B q_{2,k}^{*2} ] \,,
\end{equation}
with $A$ and $B$ constants. 
However, in order for there to be a narrow resonance, the $A$ and $B$ terms must cancel, so
that one is, in general, outside the range of convergence of the effective range expansion.

We close this subsection by noting that, within the $s$-wave approximation, $\Kdf$ becomes a function solely
of the total energy and the spectator momenta, $\Kdf = \Kdf (E, \vec p, \vec k)$. This, along with the above-described approximations, leads to
a set-up that is analogous to that 
used to date in the NREFT and FVU approaches~\cite{\Akakia,\Akakib,\MD,\Akakinum,\MDpi}.
However, 
a careful analysis of the definition of $\Kdf$, which depends implicitly on $\K_2$, reveals that it is not consistent with particle-interchange symmetry to restrict
$\Kdf$ to $s$-waves in the $\ell,m$ indices, while allowing dependence on the spectator
momenta, $\vec k$ and $\vec p$. Instead one must apply a consistent truncation across the two- and three-particle sectors. In the case that $\K_2$ is restricted to the $s$-wave, this implies that $\Kdf$ must be evaluated in the isotropic approximation, to which we now turn.

\subsection{The isotropic approximation \label{sec:isotropic}}

A systematic method for understanding the implications of particle-interchange and Lorentz
symmetry for $\Kdf$ is the threshold expansion~\cite{\BHSnum,\dwave}.
This is a low-energy expansion about the three-particle threshold, and is the analog of the the effective-range expansion 
described above for $\K_2$ [see Eq.~(\ref{eq:effrange})].
At leading order in this expansion, corresponding to keeping only the $-1/a_0$ term in  (\ref{eq:effrange}),
$\Kdf$ is a constant, which we denote $\Kiso$. 
 Here, the superscript ``iso'' stands for isotropic. Generally we take this to mean that the function is independent of (i.e.~constant with respect to) all coordinates besides the total three-particle energy.
At the next order in the threshold expansion, corresponding to the $ q_{2,k}^{*2} $ term in Eq.~(\ref{eq:effrange}), 
$\Kdf$ becomes a linear function of $E^2$ while remaining isotropic~\cite{\dwave}.
When expressed in terms of the $\vec k,\ell,m$ indices, this implies that it is pure $s$-wave,
and independent of the initial and final spectator momenta.
At quadratic order in the threshold expansion, there are isotropic terms quadratic in $E^2$,
but, in addition, terms in $\Kdf$ arise that are no longer isotropic
and contain both $s$- and $d$-waves in the $\ell, m$ indices~\cite{\dwave}.
Thus, within the context of the threshold expansion, it is only consistent to allow an 
isotropic $\Kdf$ with a  linear dependence on $E^2$.

An isotropic $\Kdf$, together with the choice $\ell_{\rm max}=0$---a combination that
we call, following Refs.~\cite{\HSQCa,\BHSnum}, the isotropic approximation---was 
shown in Ref.~\cite{\HSQCa} to reduce the $\Kdf$-dependent part of the quantization condition
(\ref{eq:QC}) to a one-dimensional algebraic equation,
\begin{equation}
F_3^{\iso}(E,L)^{-1} + \Kiso(E) = 0\,.
\label{eq:QCiso}
\end{equation}
Here $F_3^\iso$ is the isotropic projection of $F_3^s$,
\begin{equation}
F_3^\iso = \langle \mathbf{1} | F_3^s |\mathbf{1} \rangle\,,
\label{eq:F3iso}
\end{equation}
where $F_3^s$ is given in Eq.~(\ref{eq:F3s}),
and the vector $|\mathbf{1}\rangle$ has a unit entry for each
value of the spectator momentum lying below the cutoff. To reach Eq.~(\ref{eq:QCiso}), in addition to the isotropic approximation, one must project onto the trivial finite-volume irrep, denoted $\mathbb A_1^+$.\footnote{The $\mathbb A_1^+$ is only the trivial irrep for three scalar particles, while for pseudoscalars it is the $\mathbb A_1^-$.
} In previous work we indicated that (\ref{eq:QCiso}) describes all finite-volume energies that are shifted by interactions. In fact this is not the case; levels that are independent of $\Kdf$, but shifted by $\K_2$, appear in other irreps.
 
The procedure for solving the quantization condition in the isotropic approximation is simple in principle:
given $L$, $\K_2$ and $\Kiso$,  the left-hand side of (\ref{eq:QCiso}) becomes a known function of 
$E$ and the solutions can be numerically determined with a suitable root-finding algorithm.
Details of our numerical implementation are given in Ref.~\cite{\BHSnum} and are unchanged here.

As noted above, if one consistently uses the threshold expansion, then the isotropic
approximation requires truncating the effective range expansion (\ref{eq:effrange}) at second
order, and allowing only a linear dependence of $\Kiso$ on $E^2$.
For simplicity, however, in our numerical studies we mostly keep only
 the leading-order terms in the threshold expansion,
 so that interactions are described in terms of two constants, $a$ and $\Kiso$.
 We also consider the case of a constant $\Kiso$ and a Breit-Wigner form for the phase shift,
 Eq.~(\ref{eq:BW}). Though not consistent with the threshold expansion power counting, we view this as
 a reasonable starting point for studying the impact of two-particle resonances on the 
 three-particle spectrum.

\subsection{Including $d$-wave interactions: $\ell_{\rm max}=2$}

We also include in our numerical examples a study with $\ell_{\rm max}=2$,
so that $d$-wave two-particle interactions are included, as well as three additional
terms in $\Kdf$, two of which are not isotropic. A complete description of the set up has been given
in Ref.~\cite{\dwave}, and we do not repeat it here. We note only that the $d$-wave contribution to
$\wt\K_2$ has the form
\begin{equation}
{\wt \K_2^d(\vec k)} =
\frac{32 \pi \omega_k E_{2,k}^*}{
-1/(a_2^5\, q_{2,k}^{*4})  + |q_{2,k}^*| [1-H(\vec k)]}\,,
\label{eq:K2td}
\end{equation}
which should be compared to the $s$-wave form of Eq.~(\ref{eq:K2t}).
Here we are keeping only the leading non-trivial term in the $d$-wave effective range expansion, 
which is parametrized by the scattering length $a_2$. In this truncation, energies in non-trivial irreps are also shifted from their noninteracting values. Nonetheless in this work we restrict attention to the $\mathbb A_1^+$ for simplicity. See Ref.~\cite{\dwave} for interacting solutions in the two-dimensional $\mathbb E^+$ irrep.

\section{Generalizing the quantization condition \label{sec:generalizing}}

In order for the derivation of the quantization condition, Eq.~(\ref{eq:QC}), to be valid, it is necessary
that all angular-momentum components of the modified K matrix, $\wt\K_2^{(\ell)}$, have no singularities for all c.m.~frame two-particle energies, $E_{2,k}^*$, in the range $0 < E_{2,k}^* < 4m$. 
As we recall below, this implies that the quantization condition cannot be used for cases in which
there are two-particle bound states or resonances, and is a significant restriction on the applicability of the formalism.
It turns out, however, that there is a simple way to 
generalize the formalism such that, for each value of $\ell$, the inverse of $\wt\K_2^{(\ell)}$ is shifted by an arbitrary real
function of $q_{2,k}^{*2}$.  This leads to a second version of the modified K matrix in which the problematic singularities have been removed, and this freedom is
sufficient to extend the applicability of the formalism to include two-particle bound states and resonances.
In this section we explain this generalization and describe its implementation,
leaving a detailed derivation to a separate paper~\cite{inprog}.
We give three examples of how the approach may be applied. 
First, in Sec.~\ref{sec:bounds}, we consider $s$-wave bound states;
then, in Sec.~\ref{sec:boundd}, we discuss bound states in the $d$-wave;
and finally, in Sec.~\ref{sec:BW}, we turn to the case of two-particle resonances.
 
\begin{figure}[h!]
\begin{center}
\includegraphics[width=\textwidth]{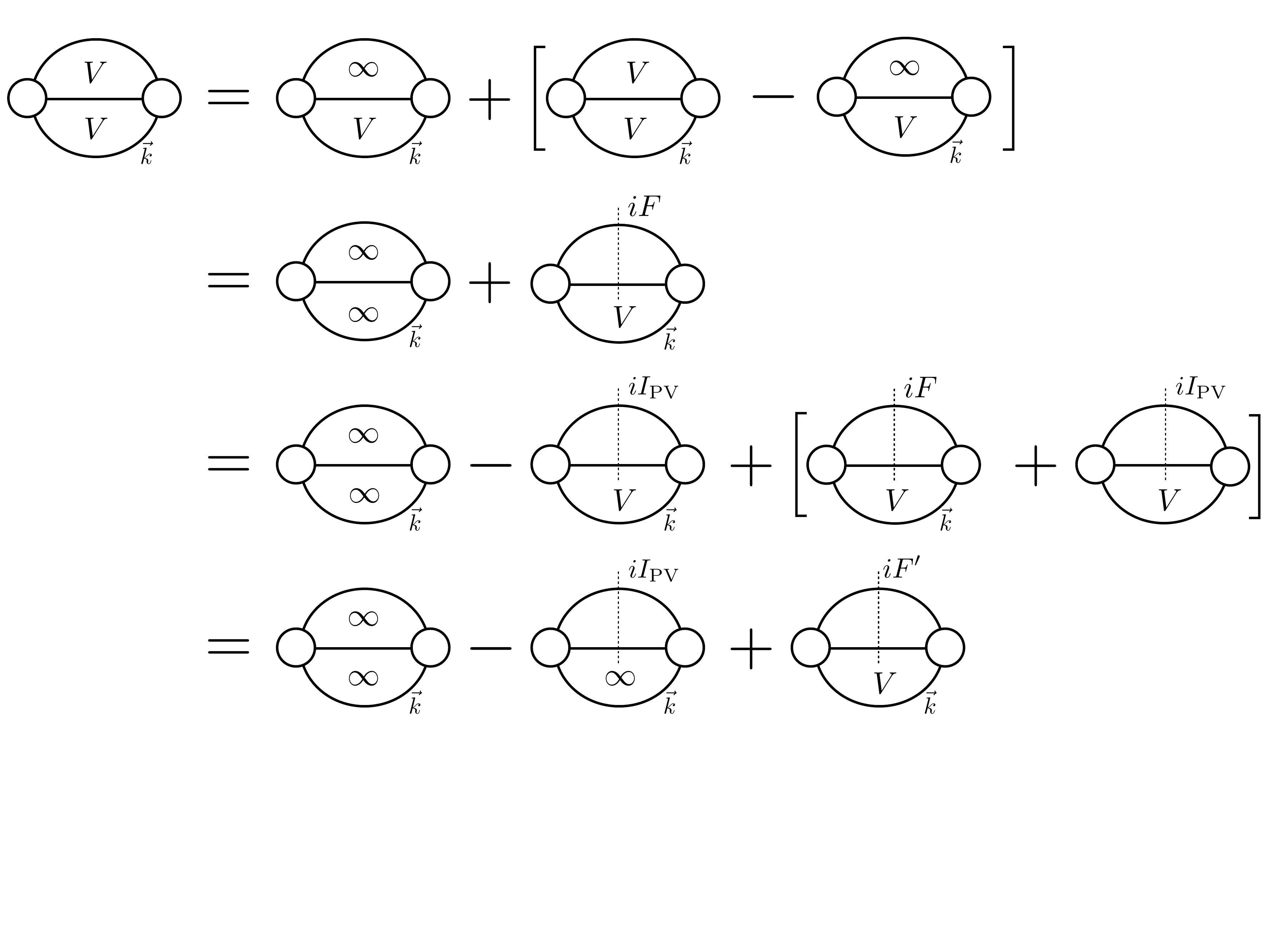}\vspace{-3.0cm}
\caption{ 
Illustration of the previous and modified PV prescriptions used in the analysis of
the three-particle correlation function, for the simplest diagram. Loops containing 
the symbol ``$\infty$'' are integrated, while those containing a ``$V$'' have the spatial components
of the momentum summed. All integrals use the PV prescription.
In the first line the upper sum is replaced by an integral plus a sum-integral difference.
The latter can be replaced by an $F$ cut, as shown in the second line, while the lower loop sum
can be replaced by an integral in the first term on this line.
In the third line we subtract and add the $I_\PV^s$ cut.
The final line shows that this new cut can be absorbed into the modified $F$,
leaving a modified PV prescription for the upper loop integral.
}
\label{fig:IPV}
\end{center}
\end{figure}

The modification of the formalism is effected by changing the principal value (PV) prescription 
used in multiple places in the derivation of Ref.~\cite{\HSQCa}. 
We recall that the original derivation consists of evaluating a finite-volume two-point correlation function, projected to kinematics for which
 three-particle states may go on shell. The correlator is expressed in terms of a skeleton expansion in which all three-particle cuts are explicitly displayed, and only the sums over momenta in these types of cuts give power-like $L$-dependence. 
 One then replaces  sums with sum-integral differences plus integrals,
 which, after extensive analysis, leads to a relation  between finite-volume energies, 
 determined by the poles in the correlator,  and infinite-volume scattering quantities. 
 The first line of Fig.~\ref{fig:IPV} shows the replacement 
 for the simplest diagram contributing to the correlation function.

For those integrals that are singular due to three-particle intermediate states,
such as the upper integral in the figure,  a pole prescription is required.
The PV prescription is used so that the result is a smooth function of the lower (spectator)
momentum, $\vec k$. This allows the second step shown in the figure to be made, in which the
lower sum is replaced by an integral (for which a pole prescription is not needed).
Thus the PV prescription appears in the sum-integral difference, $F$,
as shown by the example of the $s$-wave restriction, $\wt F^s$, given in Eq.~(\ref{eq:Fts}).
It also appears in the definition of $\wt \K_2^{(\ell)}$ and in $\Kdf$.
We note that it was found in Ref.~\cite{\HSQCa} that the integrals requiring a PV prescription
are all single-loop integrals over  a spatial momentum with the integrand having
a pole of the form shown in Eq.~(\ref{eq:Fts}).

In Ref.~\cite{inprog} (\emph{to appear}) we show that 
the derivation of the quantization condition holds for a large family
of pole prescriptions, which we denote collectively by $\PV'$.
We first describe these for $\ell_{\rm max}=0$, so that only $s$-wave quantities enter. 
If the integrand is nonsingular, then no prescription is needed, and
the $\PV'$ and $\PV$ results are the same.
If the integrand has a pole, then the modification is
\begin{multline}
\PV' \int \frac{d^3 a}{(2\pi)^3} \frac{H(\vec a) H(\vec b)}
{8 \omega_a\omega_b(E-\omega_k-\omega_a-\omega_b)}
= \\
\PV \int \frac{d^3 a}{(2\pi)^3} \frac{H(\vec a) H(\vec b)}
{8 \omega_a\omega_b(E-\omega_k-\omega_a-\omega_b)}
 -\
\frac{I_\PV^s(q_{2,k}^{*2})}{32\pi}
\,.
\label{eq:newPV}
\end{multline}
Here $I_\PV^s$ is any smooth function of $q_{2,k}^{*2}$, while the negative sign 
and the $32\pi$ are for later convenience. The modified prescription is illustrated in the final line of Fig.~\ref{fig:IPV}.
This provides a complete description of the prescription for the purposes of the derivation of
the quantization condition.\footnote{%
 If the integrand is multiplied by a function of $\vec k$ and $\vec a$
having only an $s$-wave component in the c.m.~frame of the non-spectator pair (as must be the case when setting $\ell_{\rm max}=0$),
then the $I^s_\PV$ term is multiplied by the on-shell value of this function,
obtained in the manner described in Ref.~\cite{\HSQCa}.
This result holds because the difference between the on-shell and off-shell values of this function
cancels the pole, leading to a nonsingular integral that does not require a pole prescription.}
The effect of the change in prescription 
for $\wt F^s$ and $\wt \K_2^s$ is
\begin{align}
[\wt F^s]_{kp} &\to [\wt F^s]_{kp} 
+ \delta_{kp} \frac{H(\vec k)}{2\omega_k} \frac{I_\PV^s(q_{2,k}^{*2})}{32\pi}\,,
\label{eq:Fshift}
\\
\big[ ({\wt \K_2^s})^{-1} \big]_{kp} &\to\
\big[ ({\wt \K_2^s})^{-1} \big]_{kp}
- \delta_{kp} \frac{H(\vec k)}{2\omega_k} \frac{I_\PV^s(q_{2,k}^{*2})}{32\pi}\,.
\label{eq:Kshift}
\end{align}
The shift in $\wt F^s$ follows directly from Eq.~(\ref{eq:newPV}), 
while that in $\wt \K_2^s$ can be derived by enforcing the prescription-independence of the
physical quantity $\cM_2$.
There is no change in $\wt G^s$, since it does not contain an integral.

The final quantity affected by changing the $\PV$ prescription is $\Kdf$.
The change to $\Kdf$ can be determined in principle by studying the
 infinite-volume integral equations relating $\Kdf$ to $\cM_3$.
We will describe these changes in Ref.~\cite{inprog}, and only note here three important results:
\emph{(i)} if $\Kdf$ vanishes, then it remains zero after the change in prescription;
\emph{(ii)} the changes in $\Kdf$ are, in general, energy dependent;
\emph{(iii)} an isotropic $\Kdf$ is changed, in general, into a nonisotropic one, containing partial
waves beyond $s$-wave. 

We stress that the quantization condition
containing the new versions of $\wt F$, $\wt\K_2$ and $\Kdf$,
must lead to the same relation between the physical scattering amplitudes and the finite-volume spectrum, for all $L$, up to exponentially-suppressed
effects. All we are doing is reshuffling contributions in the all-orders diagrammatic analysis of the
same finite-volume correlator.
Nevertheless, once we make an approximation, 
the spectrum need no longer be independent of $I_\PV^s$, and the result \emph{(iii)} above
shows that, indeed, the spectrum cannot be left strictly invariant if we maintain an isotropic $\Kdf$ while varying $I^s_{\PV}$.
In other words, we cannot  exactly compensate at all values of $L$ for the shifts in $\wt F^s$ and 
$\wt\K_2$ by an (energy-dependent)  shift in $\Kiso$. 
Indeed, we can use the residual $I_\PV^s$ dependence as an estimate of the error due to
truncating the quantization condition.
This is analogous to the use of scheme-dependence as an estimate of truncation errors
in quantities calculated in perturbation theory.
We also note that there is no {\em a priori} theoretical reason 
to favor any particular choice of $I_\PV^s$.

We investigate the size of the $I_\PV^s$ dependence in an example presented in Appendix~\ref{app:IPV},
finding it to be numerically small.
We also note in the appendix that we can use the threshold expansion to estimate the
size of any residual $I_\PV^s$ dependence. In particular, when we enforce the isotropic approximation,
residual $I_\PV^s$ dependence will be suppressed by
$\cO(\Delta^2)$, where $\Delta=(E^{*2}-9m^2)/(9m^2)$, since nonisotropic terms in $\Kdf$
enter only at this order in the threshold expansion~\cite{\dwave}.

The only exception to the above discussion occurs when $\Kiso$ vanishes.
The result \emph{(i)} implies that no change to $\Kiso$ is then needed to maintain the same physical scattering amplitude and spectrum
as $I_\PV^s$ is varied. Indeed, this is seen both in the integral equations relating $\Kdf$ to $\mathcal M_3$, to be discussed in Ref.~\cite{inprog}, and in the quantization condition, Eq.~(\ref{eq:QCiso}), relating $\Kdf$ to $E_n(L)$. In both cases, when $\Kiso = 0$ all $I_\PV^s$ dependence drops out.
To see this in detail in the quantization condition note that when $\Kiso=0$, a solution requires that $F_3^\iso$ diverges. 
Looking at the form of $F_3^s$, Eq.~(\ref{eq:F3s}), we see that it only diverges either if
$\wt F^s$ diverges or if the denominator of the second term,
\begin{equation}
H_{FG} =1/\wt\K_2^s + \wt F^s + \wt G^s\,,
\label{eq:HFG}
\end{equation}
has a zero eigenvalue.
The former possibility leads to poles at free three-particle energies, which are known from the
analysis of Ref.~\cite{\dwave} to be absent in the isotropic approximation, cancelling between the two terms defining $F_3^s$.
Thus the only solutions come from zero eigenvalues of $H_{FG}$.
However, as can be seen from Eqs.~(\ref{eq:Fshift}) and (\ref{eq:Kshift}),
the shifts in $\wt F^s$ and $1/\wt \K_2^s$ exactly cancel, so that $H_{FG}$ is independent of $I_\PV^s$.

The latter result means, in practice, that we do not need to introduce $I_\PV^s$
when determining solutions with $\Kiso=0$, 
and can use exactly the same numerical method
as in Ref.~\cite{\BHSnum}. 
For nonvanishing $\Kiso$ however, we do need to choose a nonvanishing $I_\PV^s$ such that
the quantization condition is valid for the chosen value of the two-body scattering parameters.
In practice, we have found it sufficient to consider only the case of a constant $I_\PV^s$.
Introducing $I_\PV^s$ into the numerical analysis is very straightforward, and
the methodology of Ref.~\cite{\BHSnum} can be carried over essentially without change.

Before turning to our three specific examples, we close the general discussion with the extension of our new $\PV'$ prescription to higher angular momenta. This is straightforward and
the essential feature is that, for each (allowed) choice of $\ell$, one has the freedom to
introduce a different real, smooth function, $I_\PV^{(\ell)}(q_{2,k}^{*2})$. 
The changes to $\wt F$ and $\wt\K_2$, which are now matrices with the full set of indices,
are 
\begin{align}
[\wt F]_{k\ell' m';p\ell m} &\to [\wt F]_{k\ell' m';p \ell m} 
+ \delta_{kp} \delta_{\ell'\ell}\delta_{m' m}
\frac{H(\vec k)}{2\omega_k} \frac{I_\PV^{(\ell)}(q_{2,k}^{*2})}{32\pi}\,,
\label{eq:Fellshift}
\\
\big [  (\wt \K_2)^{-1} \big]_{k\ell' m'; p \ell m} &\to\
\big [  (\wt \K_2)^{-1} \big]_{k\ell' m'; p \ell m} 
- \delta_{kp} \delta_{\ell'\ell}\delta_{m' m}
\frac{H(\vec k)}{2\omega_k} \frac{I_\PV^{(\ell)}(q_{2,k}^{*2})}{32\pi}\,,
\label{eq:Kellshift}
\end{align}
where $\ell$ and $\ell'$ are even.
As before, if we set $\Kdf=0$, as we do in some of the numerical examples below,
then we can in practice ignore the $I_\PV^{(\ell)}$ shift, since it cancels
in the quantization condition. 

Further details, as well as a discussion of the relationship between the introduction of
$I^{(\ell)}_\PV$ and the formalism presented in Refs.~\cite{\BHSQC,\BHSK}, where the K matrix poles were taken into account explicitly, will be given in Ref.~\cite{inprog}.

\subsection{Using $I_\PV^s$ to accommodate an $s$-wave bound state \label{sec:bounds}}

Here we show that, if we set $I_\PV^s$ to an appropriate constant, then the quantization
condition in the $s$-wave, isotropic approximation is valid
for three-particle systems in which there is a two-particle scalar bound state.
Specifically, we consider the case in which we 
keep only the leading term in the ERE, i.e.~the scattering length
in Eq.~(\ref{eq:effrange}). This is also one of the examples that we 
investigate numerically below.

\begin{figure}[H]
\centering 
\subfigure[ \label{fig:k2inv3} $I_{\PV}^s=0$]{\includegraphics[width=.49\textwidth]{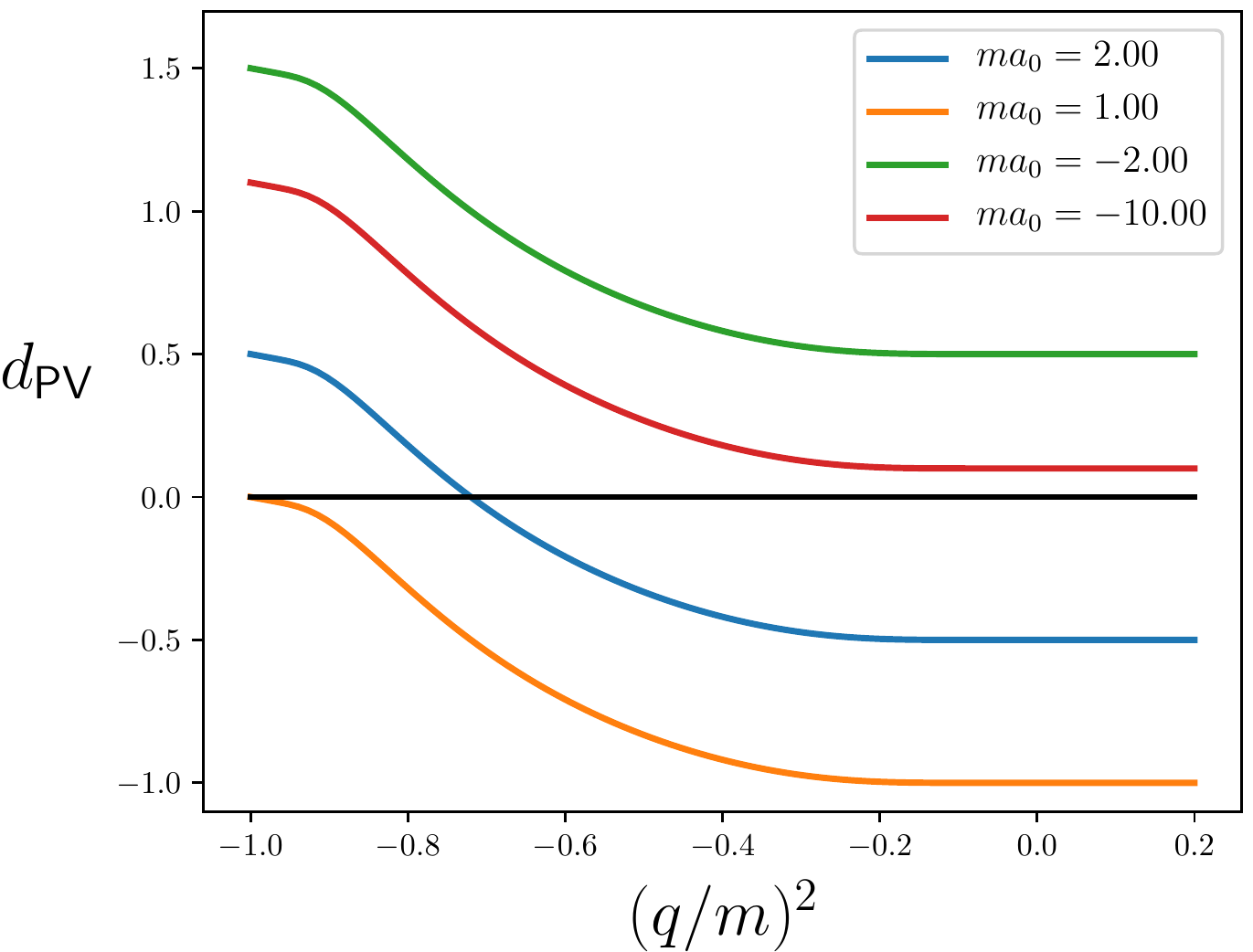}}
\hfill
\subfigure[ \label{fig:k2inv4} $I_{\PV}^s=-1$]{\includegraphics[width=.49\textwidth]{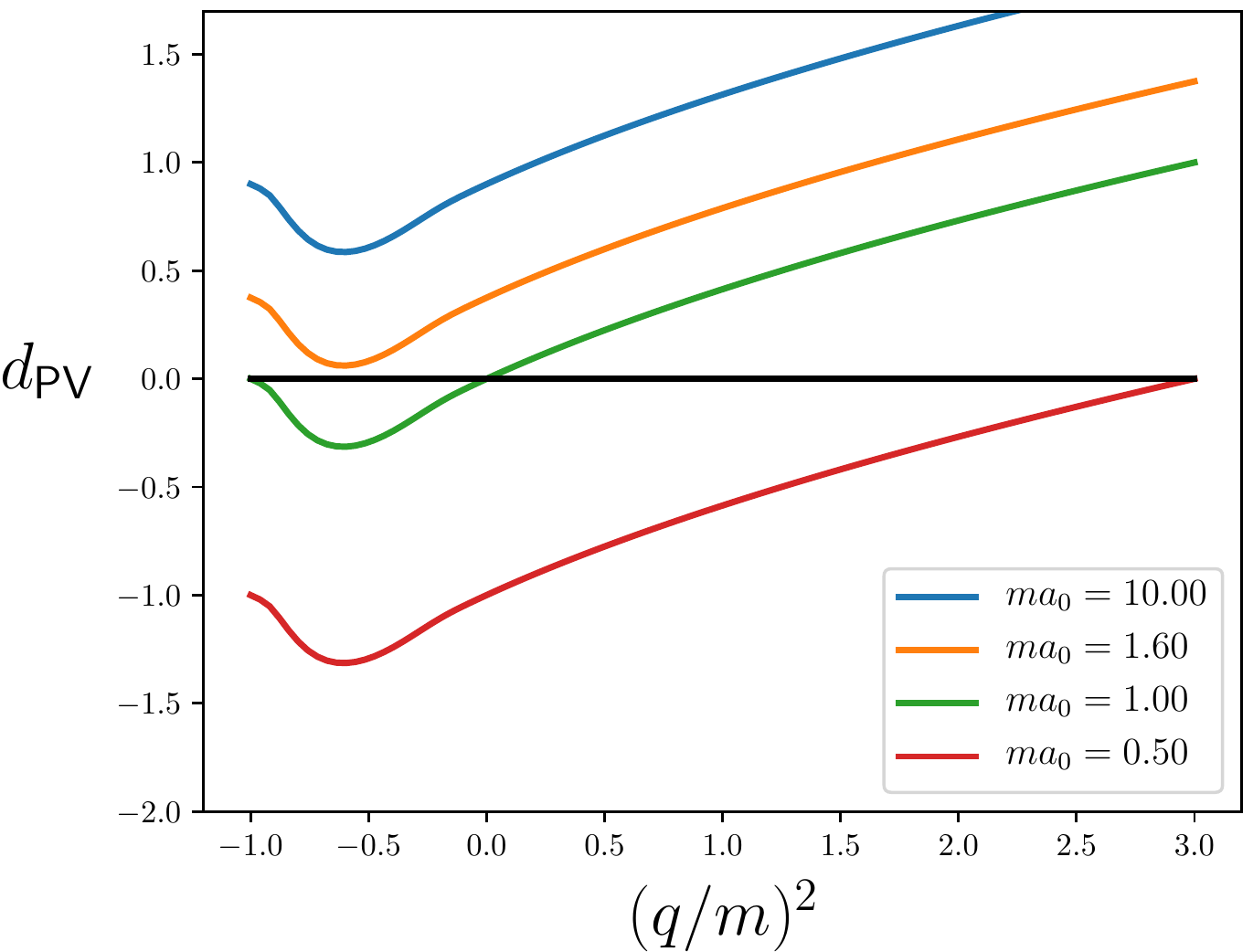}}
\hfill
\caption{Plots of the denominator of $\wt \K_2^s$, $d_\PV(q^2)$,
 vs. $q^2/m^2$ for a range of choices of $am$ with $I_{\PV}^s=0$ (left) 
and $I^s_{\PV}=-1$ (right). 
There is a pole in $\wt \K_2^s$ whenever $d_\PV(q^2)$ vanishes.
See text for further discussion.
\label{fig:k2inv34}}
\end{figure}

The quantization condition is valid as long as $\wt \K_2^s$ has no pole in the
kinematic range of interest. To study this, we consider the denominator of $\wt \K_2^s$,
which, from Eq.~(\ref{eq:K2t}), together with the modification given in Eq.~(\ref{eq:Kshift}),
is\footnote{%
Previously we have written $H$ as a function of the spectator momentum, $\vec k$.
The explicit form that we use, given in Ref.~\cite{\BHSnum}, is in fact a function of $q_{2,k}^{*2}$
(itself a function of $\vec k$) and it is more convenient here to make this explicit,
at the cost of some abuse of notation.}
\begin{align}
d_\PV(q^2) & \equiv \frac{32 \pi \omega_k E_{2,k}^*}{\wt \K_2^s} \,, \\ 
&  = -1/(m a_0) +  [1- H(q^2)] |q|/m - I_\PV^s H(q^2) \sqrt{1+q^2/m^2}
\,.
\label{eq:denKs}
\end{align}
There is a pole whenever this quantity vanishes.
We plot $d_\PV(q^2)$ for $I_\PV^s=0$ and $-1$ in Fig.~\ref{fig:k2inv34}.
The cutoff function vanishes when $q^2/m^2 < -1$ 
so the lower limit in all the plots is set to this value. 
The upper limit depends on $E/m$: for $E/m=3$ it is $q^2/m^2=0$, while for
$E/m=5$ (the maximum value for which our quantization condition holds) 
it is $q^2/m^2=3$. In the left-hand figure, where $I_\PV^s=0$, the curves are flat for
$q^2/m^2> 0$, and so we do not show the entire range.

From the left panel we see that, for $I_\PV^s=0$, $d_\PV(q^2)$ has a zero-crossing when $m a_0 > 1$,
a result that is simple to verify analytically.
We stress that the pole that appears lies far below threshold.
For example, as $ma_0 \to 1^+$  the pole position approaches $q^2/m^2= -1$.
The pole is not related to a physical quantity, as is made clear by the fact that its position
depends on the cutoff function $H(\vec k)$.
Nevertheless, it presents a barrier to the derivation of the quantization condition.

The restriction to $ma_0 \le 1$ implies that the formalism cannot accommodate 
a two-particle bound state when using $I_\PV^s=0$.
To understand this, recall that a bound state occurs when $\cM_2$ has a pole for $q^2 < 0$,
i.e.~when $q \cot\delta_s(q)+|q| =0$. 
In our approximation, this becomes $q^2 = -1/a_0^2$.
Since our cutoff function leads to the restriction $q^2 > -m^2$,
the bound state is present only for $ma_0 > 1$.  

The right panel, Fig.~\ref{fig:k2inv4}, shows that, using the $\text{PV}'$ prescription
with $I_\PV^s=-1$, the quantization condition is now valid for $ m a_0 < 0.5$ and $m a_0 \gtrsim 1.6$.
This shift in the range of validity continues as $I_\PV^s$ is lowered further.
For $I^s_\PV=-2$ the range becomes $m a_0< 0.25$ and $m a_0 \gtrsim 1.15$.
This raises the question of whether, for any choice of $m a_0 > 1$, there is a 
value, or range of values, of $I^s_\PV$ for which the quantization condition is valid. 
We find numerically that the answer is affirmative---for a given choice of $m a_0 >1$, as long
as $I_\PV^s$ is sufficiently negative, the quantization condition holds. 
The minimum value of $I_\PV^s$ that is needed grows rapidly as $m a_0$ approaches unity.
Nevertheless, the key point is that we can use the quantization condition to study all values
of the scattering length by choosing $I_\PV^s$ to lie in an appropriate range
that depends on the value of $m a_0$.

\subsection{Using $I^d_\PV(q^2)$ to accommodate a $d$-wave bound state\label{sec:boundd}}

We now turn to the case that a pole appears in the $\ell=2$ component of our K-matrix-like quantity. In this case, the denominator of $\wt \K_2^d$, Eq.~(\ref{eq:K2td}), is modified by
Eq.~(\ref{eq:Kellshift}) to
\begin{align}
d^d_\PV(q^2) & \equiv \frac{32 \pi \omega_k E_{2,k}^*}{\wt \K_2^d} \,, \\ 
& = -1/(a_2^5 q^4m) +  [1- H(q^2)] |q|/m - I^d_\PV(q^2) H(q^2) \sqrt{1+q^2/m^2}
\,.
\label{eq:denKd}
\end{align}
This differs from $d^s_\PV(q^2)$, Eq.~(\ref{eq:denKs}), 
only by the presence of the $1/q^4$ factor in the first term on the right-hand side.
The quantization condition is valid as long as $d^d_\PV(q^2)$ does not vanish for $q^2$ in the
allowed kinematical range, $-1 < q^2/m^2 < 3$.
It is straightforward to show that, for $I^d_\PV=0$, such a zero crossing only occurs 
when the $d$-wave scattering length satisfies $m a_2  > 1$.
This is also the condition for a $d$-wave bound state to be present:
\begin{equation}
-\frac1{a_2^5 q^4} + |q| = 0 \ \ \Rightarrow\ \ |q| = \frac1{a_2}  \,,
\end{equation}
which, since $|q| <  m$, implies that one must have $m a_2  > 1$.
Thus previous numerical investigation of the quantization condition
including $d$-waves was restricted to values of the scattering length such
that there were no $d$-wave dimers.

This restriction can be lifted using the $\PV'$ prescription.
A simple choice is to set $I^d_\PV(q^2)= c/q^4$, with $c$ a constant. Then, by multiplying
$d^d_\PV(q^2)$ by $q^4$, the analysis becomes very similar to that for $d^s_\PV(q^2)$, and
one finds that, for any value of $m a_2 $, there is a range of values of $c$ for which $d_\PV^d(q^2)$
has no zero crossing. We use this freedom in the numerical investigations described
in Sec.~\ref{sec:numdwave}.

\subsection{Using $I^s_\PV$ to accommodate an $s$-wave resonance}
\label{sec:BW}

\begin{figure}[H]
\centering 
\subfigure[ \label{fig:K2inv0} $I^s_{PV}=0$]{\includegraphics[width=.49\textwidth]{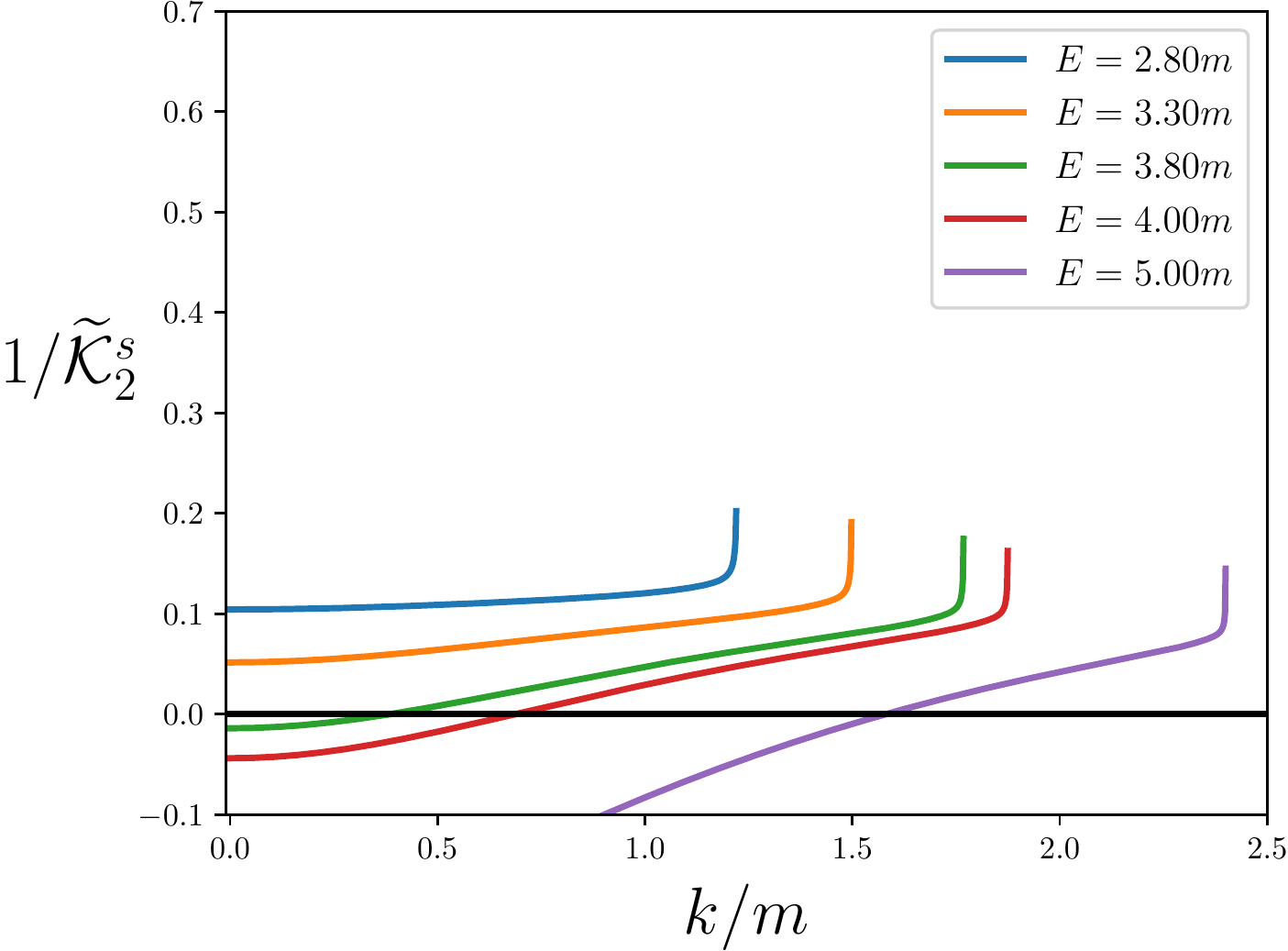}}
\hfill
\subfigure[ \label{fig:K2inv100} $I_{PV}=-100$]{\includegraphics[width=.49\textwidth]{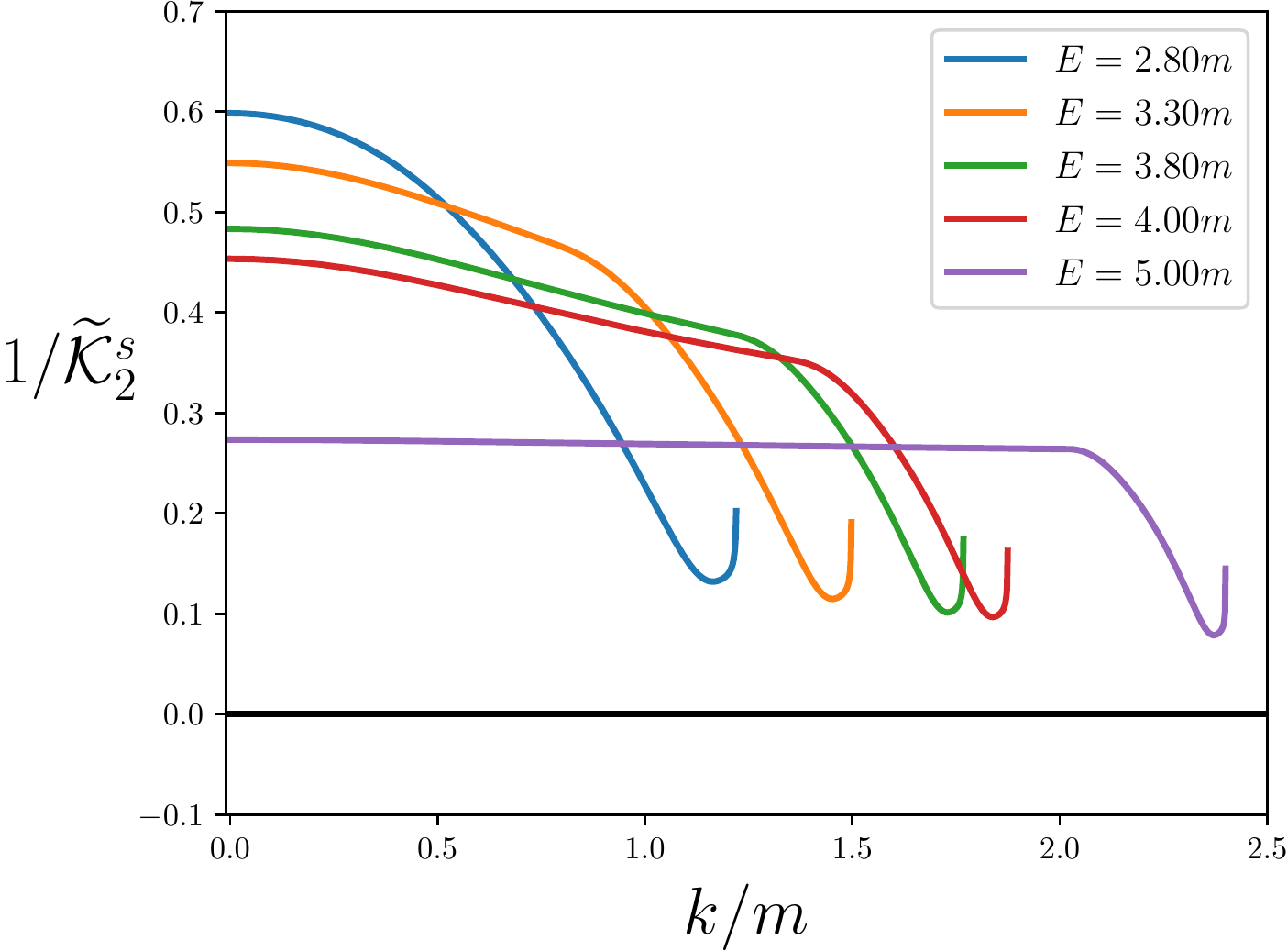}}
\hfill
\caption{\label{fig:K2inv} 
$1/\wt\K_2^s$ vs $|\vec k|/m$, with $\vec k$ the spectator momentum for
the Breit-Wigner form of the phase shift given in Eq.~(\ref{eq:BW}),
with the resonance parameters set to $g=1$ and $m_R=2.7 m$.
Thus the resonance only leads to a  zero crossing for  $E > 3.7 m$.
Results are shown for (a) $I_\PV^s=0$ and (b) $I_\PV^s= -100$.
}
\end{figure}

Our final example of using the $\text{PV}'$ prescription is for an $s$-wave resonance with the
phase shift given in Eq.~(\ref{eq:BW}). Here we find it more convenient to show plots
of $1/\wt \K_2^s$ vs.~$|\vec k|$, and these are shown in Fig.~\ref{fig:K2inv}
for the choice of resonance parameters used in our numerical investigations below.
Figure~\ref{fig:K2inv0} shows the result with the original PV prescription, showing the zero associated
with the resonance, and its dependence on $E$.
The right panel, Fig.~\ref{fig:K2inv100}, shows that by choosing the constant value $I_\PV^s=-100$,
the zero can be removed for the kinematic range of interest.

\section{Numerical results\label{sec:numerical}}

In this section we present numerical results from the quantization condition
using the $\PV'$ prescription, which allows us to consider choices for $\K_2$ that
were inaccessible to our previous numerical explorations~\cite{\BHSnum,\dwave}.
Our approach is to assume forms for $\K_2$ and $\Kdf$
and determine the resulting finite-volume spectrum. In a practical application, using LQCD finite-volume energies, one will ultimately identify a broad set of $\K_2$ and $\Kdf$ parametrizations and fit these to the numerically-determined spectrum, ideally for various values of $L$ and various total spatial momenta. This idea, proposed in Ref.~\cite{Guo:2012hv} and now standard in the analysis of two particle channels, allows one to identify the subset of parametrizations that can describe the physical system under consideration. In addition, the spread in the functional forms $\K_2$ and $\Kdf$ that give a good description provides a systematic uncertainty, indicating how well the input finite-volume information can constrain these infinite-volume objects.
Here, instead, we aim to illustrate the types of behavior that can be expected in the spectrum for different fixed
choices of the two- and three-particle interactions.

We begin in Sec.~\ref{sec:largea0} by working in the isotropic approximation, presenting
a global view of the spectrum for values of the $s$-wave scattering length, $ma_0 > 1$, 
such that there are two-particle bound states (called dimers).
We consider a range of choices of $a_0$, including those in which the dimer is deeply bound, 
requiring a relativistic formalism such as ours,
as well as those for which the bound state is shallow, allowing comparison
with results from the NREFT  three-particle formalism~\cite{\Akakinum}.
A feature of most of these spectra is the appearance of one or more three-particle bound states 
(called trimers).

Next, in Sec.~\ref{sec:dimerpart}, we focus on the region of the spectrum below the three-particle
threshold, where finite-volume states are dominantly composed of a dimer together with a particle.
Here, by going to large volumes,
 we are able to use our formalism as a tool for determining the properties of dimer-particle
scattering in infinite-volume.
This leads us, in Sec.~\ref{sec:nD},
to adjust the parameters $a_0$ and $\Kdf$ so that we can model the
three-nucleon system with deuteron and triton bound states, albeit without including spin.
This is the only example in which we consider nonvanishing $\Kdf$.

Still working in the isotropic approximation, in Sec.~\ref{sec:BWres} we determine
the form of the three-particle spectrum in the presence of a narrow $s$-wave resonance.
To our knowledge, this is the first example of such a study.
Finally, in Sec.~\ref{sec:numdwave}, we turn on $d$-wave interactions, leading to the possibility
of both $s$- and $d$-wave dimers, as well as trimers.

\subsection{Spectrum with $m a_0 > 1$}
\label{sec:largea0}

In this subsection we work in the isotropic approximation, described in Sec.~\ref{sec:isotropic},
and keep only the leading term in the effective range expansion, so that
\begin{equation}
q \cot \delta_0 = -{1}/{a_0}\,.
\end{equation}
As explained above, the key change introduced by working with $m a_0 > 1$ 
is the presence of a two-particle bound state. 
The infinite-volume mass of this dimer is given exactly by
\begin{equation}
M_d = 2m \sqrt{1 - 1/(m a_0)^2}\,, 
\label{eq:dimerswave}
\end{equation}
which varies in the range  $0 < M_d < 2m$ as $m a_0$ changes from $1$ to infinity.
At the lower end of this range, the dimer is very deeply bound and thus the internal degrees of freedom are relativistic. As $m a_0$ increases, the binding energy decreases and the bound state dynamics becomes increasingly nonrelativistic (NR). We 
expect the crossover point to be around $|q^2|/m^2 = 1/(m a_0)^2 \approx 0.1$, which
 occurs when $m a_0 \approx 3.2, \ M_d\approx 1.9m$. 
In the NR regime, the dimer wavefunction falls exponentially with a distance scale given by $a_0$. 
Thus to avoid large finite-volume effects in the dimer mass we need to use volumes such that
$a_0/L= (ma_0)/(mL) \gg 1$; in practice we require $a_0/L \gtrsim 5$. We stress that this constraint is only relevant if we wish to suppress the dimer's volume-dependence and that the quantization condition itself is valid for all choices of $a_0/L$ provided that $m L$ is large enough to safely ignore the the neglected $e^{- m L}$ scaling.

For some range of parameters we also expect there to be one or more three-particle bound states. 
In particular, we know that when $|m a_0|\gg 1$ so that we are close to the unitary limit,
Efimov trimers will form. In fact, we find trimers as soon as $m a_0$ exceeds $\approx 1.4 $. 
All calculations in this subsection are for $\Kiso=0$ so that the value of $I_\PV^s$ is irrelevant,
as explained above.

\begin{figure}[tb!]
\centering 
\includegraphics[width=1.0\textwidth]{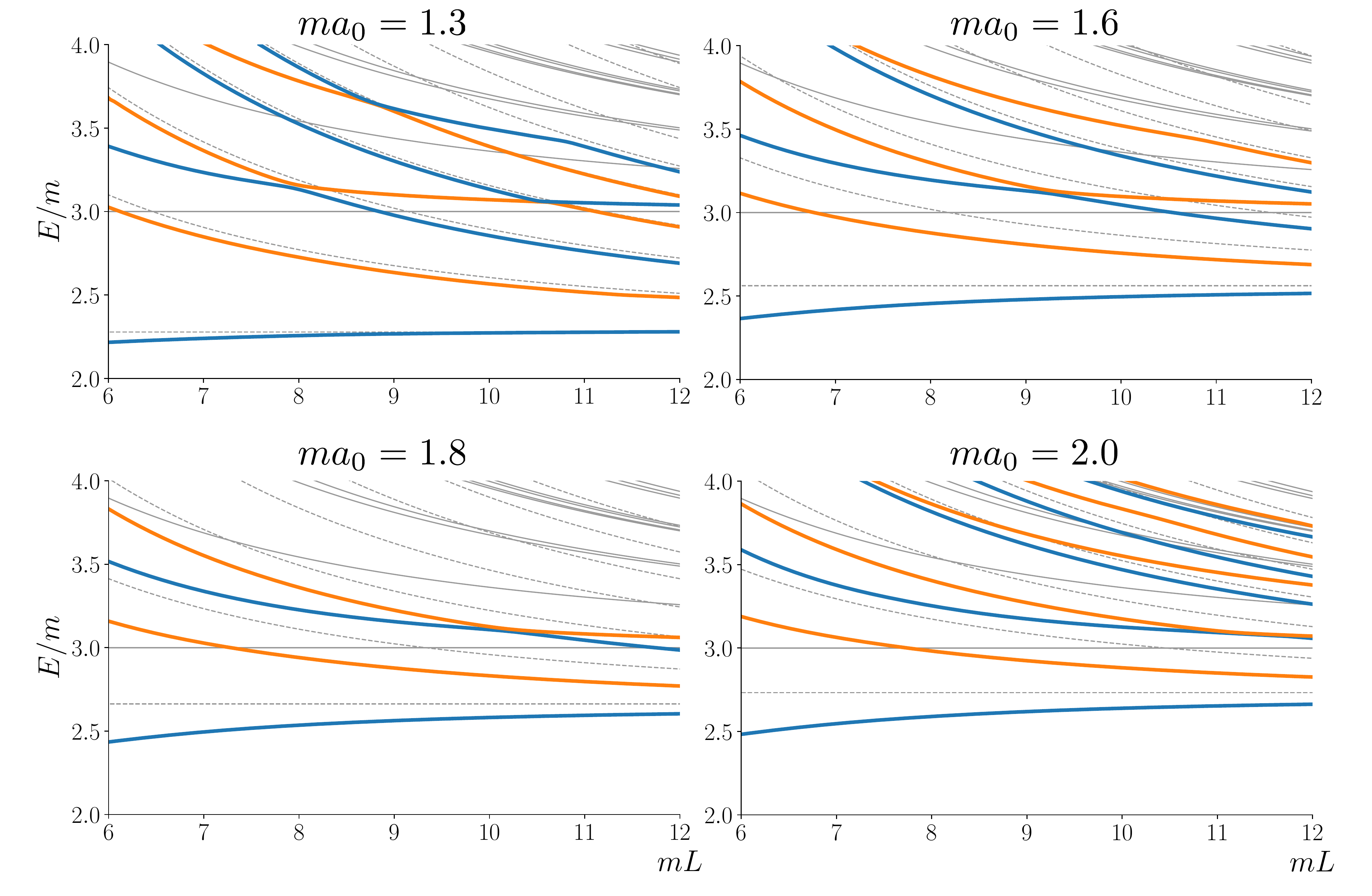}
\caption{\label{fig:spectra} 
The spectrum for four values of $ma_0$ and $\Kiso=0$ as a function of $mL$.
All four values lead to relativistic two-particle bound states.
Solid grey lines show the energies of three noninteracting particles (1$+$1$+$1 states),
while dashed grey lines give the energies of  noninteracting dimer $+$
 particle states (2$+$1 states). The interacting energy levels are shown in alternating colors to emphasize the avoided level
 crossings. Although it is not apparent from these plots, by going to larger values of $mL$
 we find that the lowest state for $ma_0=1.6$, $1.8$, and $2.0$ is a trimer, while that for $m a_0=1.3$
 asymptotes to the dimer $+$ particle energy as $mL\to \infty$. See the text for further discussion.
}
\end{figure}

\bigskip
We first determine the spectra for $1 < m a_0 \le 2$ for moderate values of $mL$,
aiming for an overview of the phenomena that can occur.
Four examples are shown in Fig.~\ref{fig:spectra}. 
To interpret the resulting spectra (shown by the colored lines in the plot) it is useful to
compare to two types of noninteracting energies.
First, there are the energies of three noninteracting particles of mass $m$, which we
refer to as $1+1+1$ levels. These are the same for all four plots. The lowest such level,
at $E/m=3$, is independent of $L$, while higher levels have $L$-dependence and asymptote to
$E/m=3$ as $L \to \infty$. The $1+1+1$ levels are shown as solid grey lines.
The second class of noninteracting energies, shown by dashed grey lines,
are dimer $+$ particle states, whose energies are given by
\begin{equation}
E_{\vec n} = \sqrt{M_d^2+ (2\pi/L)^2 \vec n^2} + \sqrt{m^2 + (2\pi/L)^2 \vec n^2}\,,
\end{equation}
with $\vec n$ an integer vector.
We refer to these as $2+1$ states for brevity.
The lowest such state has $E=M_d+m$, and all the others asymptote to this energy as
$L\to\infty$.
We note that, when we project onto the $\mathbb A_1^+$ irrep of the cubic group then each noninteracting level, both the 1+1+1 and 2+1 types, has one corresponding solution in the quantization condition.
We also stress that the dimer is a relativistic bound state for all four values of $m a_0$ shown in the figure.

\begin{figure}[tb!]
\centering 
\includegraphics[width=0.8\textwidth]{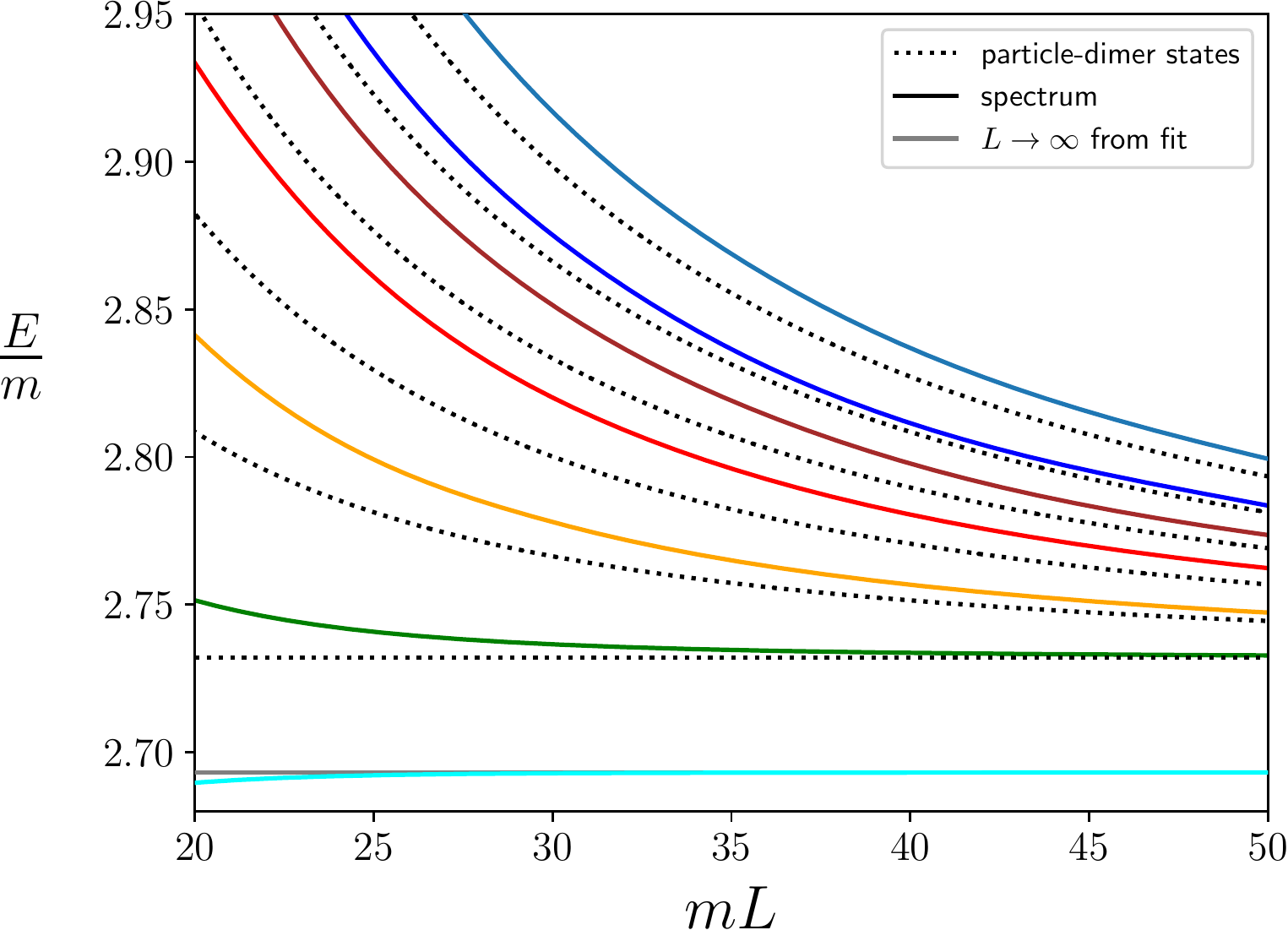}
\caption{\label{fig:am2largeL} {
Low-lying spectrum for $m a_0=2$ and $\Kiso=0$ for larger $mL$.
This is the continuation to larger volumes
of the spectrum shown in the bottom right-hand panel of Fig.~\ref{fig:spectra}.
Note the highly compressed vertical scale.
The horizontal line for the lowest level gives the $L\to\infty$ asymptote for the trimer energy,
using a fit described in the text.}
}
\end{figure}

{
The interpretation of the spectrum is simplest for $m a_0=1.3$. The lowest two levels correspond
to $2+1$ states with energies shifted down slightly by the dimer-particle interactions.
By contrast, the third level changes its nature for $mL\sim 8$: above this it is a (shifted) $2+1$ state,
while below it is a (shifted) $1+1+1$ state. This shifted state also appears in the second orange-colored
level for $8 \lesssim mL \lesssim 10$, and in the third blue level for $10 \lesssim mL$.
A similar pattern occurs for higher levels.

Although the spectrum looks superficially similar for the other (larger) values of $m a_0$,
there is, in fact, a qualitative difference. This is because, for $m a_0 \gtrsim 1.4$, a trimer appears.
The lowest (blue) level asymptotes to an energy below $M_d+m$. This is not apparent from 
Fig.~\ref{fig:spectra}, but can be seen for $m a_0=2$ by the spectrum at larger $L$ shown in
Fig.~\ref{fig:am2largeL}. Thus the interpretation of the levels for $m a_0=1.6$, $1.8$ and $2.0$ is
as follows: the lowest (blue) level is a trimer or $3$ state; the next level (orange)
is the lowest $2+1$ state with energy raised by residual interactions.
The third level begins at small $mL$ as a shifted $1+1+1$ state, but, for a value of $mL$ that depends on $ma_0$,
it changes its dominant nature to an excited $2+1$ state.
}

{
Having summarized the content of Fig.~\ref{fig:spectra}, it is instructive to consider 
Fig.~\ref{fig:am2largeL} in more detail. 
Here we restrict attention to a small energy range around the dimer $+$ particle threshold.
There are nevertheless several levels, since we work at large $mL$.
All except the trimer asymptote to $E_d+m$ as $L\to\infty$, and in the regime shown, where they
lie well below the $1+1+1$ threshold, all can be considered as dominantly $2+1$ levels,
with a repulsive interaction pushing the energies up from their noninteracting values.
In the next subsection we will do a quantitative analysis of these energy shifts, which encode information
about the dimer-particle phase shift.
For now we focus on the volume-dependence of the trimer energy, $E_t(L)$. 
We fit for $mL>25$ to the following form:
\begin{equation}
\frac{E_t(L)}{m} = \frac{E_0}{m}- \frac{|C|}{(mL)^{3/2}} e^{-2\kappa L/\sqrt3} \,,
\label{eq:trimerFV}
\end{equation}
where $\kappa^2/m^2 = 3 - E_0/m$ with $E_0=E_t(\infty)$, and $|C|$ a fit parameter.
The fit determines the asymptote to be $E_0 \approx 2.6931 m$. 
Equation~(\ref{eq:trimerFV})  is the result for the asymptotic
volume-dependence derived in Ref.~\cite{\AkakiBS} 
for a nonrelativistic bound state in the unitary (large $|m a_0|$) limit.
While it  does not obviously apply here (since $m a_0=2$ is not in the unitary regime), we
find that it gives a very good description of our results. 
However, as for the higher levels, a more rigorous approach is available for analyzing $E_t(\infty)$,
as we discuss in the next subsection.
}

\begin{figure}[H]
\begin{center}
\includegraphics[width=.8\textwidth]{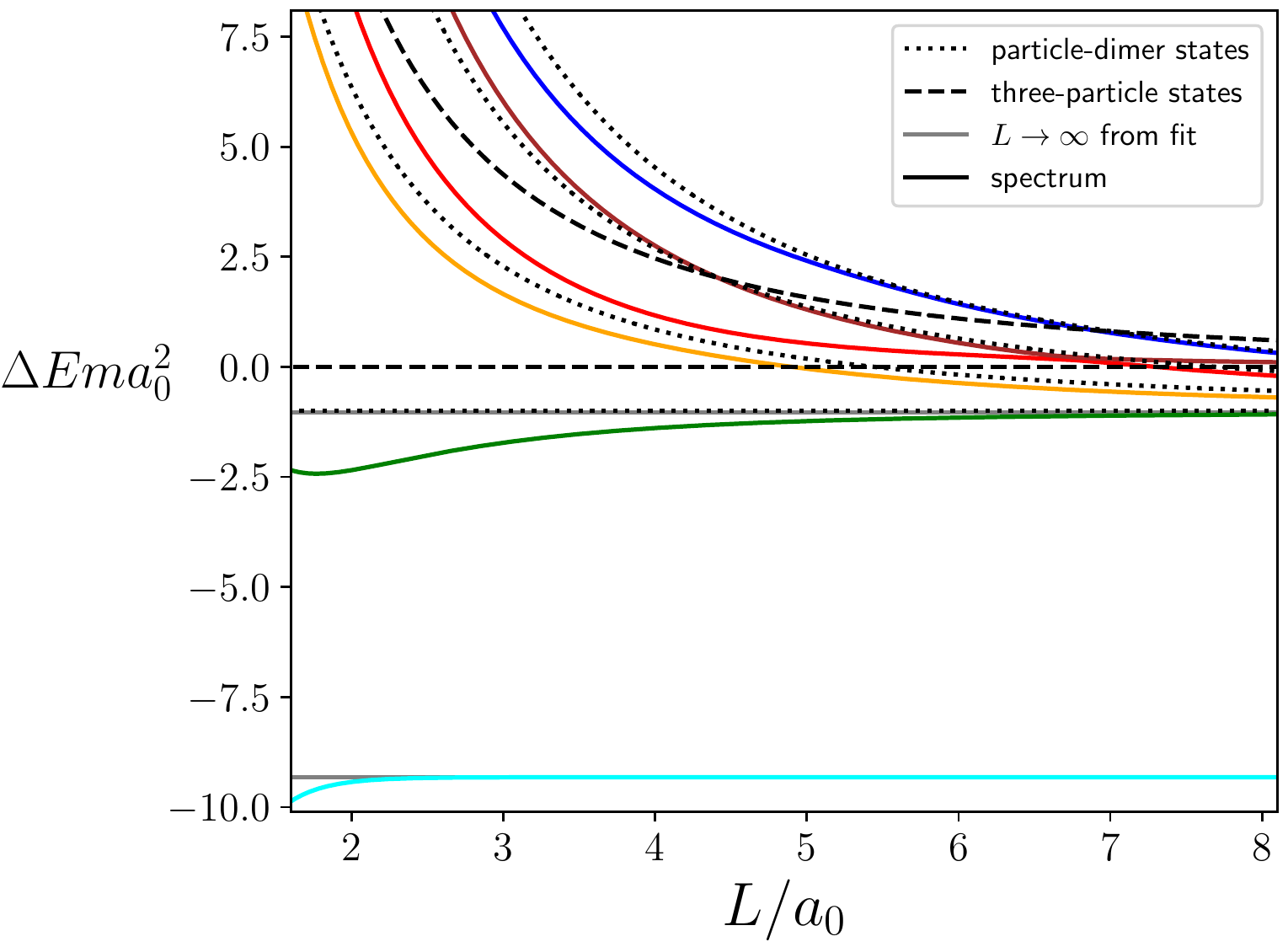}
\caption{
 Spectrum as a function of the box size for $m a_0=16$ and $\Kiso=0$, with $\Delta E=E-3m$
the nonrelativistic energy. In order to facilitate comparison with Ref.~\cite{\Akakinum},
the quantities plotted are made dimensionless using appropriate factors of $a_0$,
and thus differ from those in Figs.~\ref{fig:spectra} and \ref{fig:am2largeL}.}
\label{fig:spectrum16}
\end{center}
\end{figure}

{
We conclude our overview by studying the spectrum for $m a_0 = 16$. 
For this value of the scattering length the dimer mass lies well in the NR regime,
\begin{equation}
\frac{M_d}{m}  = 1.9961 \approx 2 -\frac{1}{(m a_0)^2}\,,
\end{equation}
so that our results can be compared to those obtained from the NREFT quantization condition,
as studied in Ref.~\cite{\Akakinum}. 
We thus display the spectrum in Fig.~\ref{fig:spectrum16}
using the variables adopted in Ref.~\cite{\Akakinum}. 
This should be compared to Figs.~3 and 6 of that work, from which the strong similarities are evident.
In particular, there are two trimers in both cases, dubbed 
the deep and shallow bound states in Ref.~\cite{\Akakinum}. 
In that work, these two trimers have energies
$\Delta E m a_0^2=(E-3m)ma_0^2=-10$ and $-1.016$, respectively.
We are not aiming to reproduce these numbers precisely, which would require tuning $\Kdf$
to nonzero values, but rather to obtain semiquantitative agreement.

To obtain the trimer energies from our results, 
we fit the lowest two spectral levels to their asymptotic forms.
For the deep (lowest) level we use Eq.~(\ref{eq:trimerFV}), obtaining
$\Delta E_t m a_0^2 = -9.3218$ 
(corresponding in relativistic units to $E/m=2.9636$).
For the shallow trimer, we treat the state as a bound state of the particle plus dimer, an
interpretation that is justified in the following subsection. Thus, following Ref.~\cite{\Akakinum}, 
we can use L\"uscher's two-particle
quantization condition to predict its asymptotic volume-dependence~\cite{Luscher:1986n1}:
\begin{equation}
\frac{E_t(L)}{m} = \frac{E_0}{m} - \frac{|D|}{m L} e^{-\kappa' L}\,,
\quad
\kappa' = \sqrt{2 \mu (3 m - E)}
\,,
\label{eq:dimerFV}
\end{equation}
with the reduced mass given by $1/\mu=1/m+1/M_d \approx 3/(2m)$.
This fit yields $\Delta E m a_0^2=-1.0301$ (corresponding to $E=2.99598 m$).
This lies below the particle-dimer threshold, given by
\begin{equation}
\Delta E_d \equiv3m-(m+M_d)=2m-M_d \approx \frac1{m a_0^2}\,,
\end{equation}
which corresponds to $\Delta E m a_0^2 = -1$ (or, strictly speaking,
$\Delta E m a_0^2 = -1.001$ if one includes relativistic effects).
The asymptotic energies are shown by the solid grey horizontal lines in Fig.~\ref{fig:spectrum16},
and the presence of the shallow dimer is indicated by the small offset between the horizontal
dashed and solid grey lines near $\Delta E m a_0^2=-1$.

A final noteworthy point of similarity between the results in Fig.~\ref{fig:spectrum16} and
those in Ref.~\cite{\Akakinum} concerns the third level (in orange) for $L/a_0\gtrsim 5$
(so that $\Delta E < 0$). This $2+1$ state lies close to the first excited
noninteracting particle-dimer energy, and far from the lowest such energy at $\Delta E m a_0^2=-1$.
Thus it appears that the latter state is missing in the spectrum. This point was observed in
Ref.~\cite{\Akakinum}, where it was argued that the missing state transmutes in finite volume into
the shallow dimer. We give further evidence for this interpretation in the following section.

\subsection{Dimer-particle scattering \label{sec:dimerpart}}

As seen in the previous subsection, states that lie below the three-particle threshold at $E=3m$ can be
interpreted as dimer $+$ particle states, abbreviated as $2+1$ states. 
In this section, we focus on this energy regime and extend our calculations to very large
$mL$, so as to learn about {\em infinite-volume} dimer-particle scattering.
In particular, we choose $L \gg a_0$, so as to avoid large finite-volume effects in the dimer,
which we know from the two-particle quantization condition fall as $\exp(- \kappa L) = \exp(-L/a_0)$.
Then, to the extent that the finite-volume states can be described as purely dimer+particle states,
we can use the nondegenerate, nonidentical form of the two-particle quantization condition, truncated to
the $s$-wave, to determine the dimer-particle scattering phase shift.
In effect, we are using the three-particle quantization condition as a tool both to solve the
relativistic two-particle bound state equation  and to determine the structure of the resulting
dimer by probing it with a third particle.
We carry out this calculation in detail for three choices of the two-particle scattering length,
$m a_0 = 2$, $6$, and $16$, again using the isotropic approximation with $\Kiso=0$.
This study extends the idea presented in Ref. \cite{Pang:2019dfe} in the nonrelativistic limit, where the scattering length of the particle$+$dimer system was related to the three-body scattering amplitude.
{
The two-particle quantization condition for non-identical scalar particles, truncated to the s-wave,
is~\cite{Luscher:1986n2,Beane:2003yx}
\begin{equation}
k \cot \delta_0(k) = \frac{1}{\pi L} \mathcal{Z}\left( \frac{L k}{2\pi} \right), 
\ \ \ \ \mathcal{Z}(\eta) =  \sum_{\vec{j}}^{ \text{UV}} \frac{1}{|\vec j|^2 - \eta^2} \,,
\label{eq:QC2}
\end{equation}
where UV indicates a suitable UV regulator, and $k$ is defined through
\begin{equation}
E = \sqrt{m^2 + k^2} + \sqrt{M_d^2 + k^2 }\,,
\end{equation}
with $M_d$ the dimer mass and $E$ the energy of the finite-volume state.}
Using Eq.~(\ref{eq:QC2}) we obtain the usual one-to-one relation between the spectral levels and
the phase shift. It is important to note that this equation holds for all levels in the spectrum
that lie in the $2+1$ regime, and not only for the lowest state.

Once we have determined $\delta_0(k)$, we use two forms to parametrize it.
The first is the standard effective range expansion (ERE),
\begin{equation}
k \cot \delta_0(k) = -\frac{1}{b_0} + \frac{1}{2} r k^2 +  P r^3 k^4 + O(k^6)\,,
\label{eq:ERE}
\end{equation}
where $b_0$ is the dimer-particle scattering length.
The radius of convergence is typically determined by the branch-point of the $t$-channel cut, or else a nearby pole in $k \cot \delta_0(k)$, corresponding to a nearby zero in the K matrix.
In the case of the latter it is helpful to use an alternative parametrization, 
taken from infinite-volume studies of nucleon-deuteron scattering~\cite{Black,rusos},
\begin{equation}
k \cot \delta_0(k) = \frac{-\frac{1}{b_0} + \frac{1}{2} r k^2 +  P r^3 k^4}{1-\frac{k^2}{\kappa^2}} + O(k^6)\,.
\label{eq:EREpole}
\end{equation}
We expect that the inclusion of the pole will increase the range over which this form provides a
good description of the phase shift.

In many cases we encounter bound states of the dimer-particle system, 
which occur whenever $k \cot \delta_0(k)=-|k|$.
For these, it is important to keep in mind the following consistency check that holds
for physical bound states~\cite{Iritani:2017rlk}, 
\begin{equation}
\frac{d}{dk^2}\left[ k \cot \delta_0(k)  - \left( - \sqrt{-k^2}  \right) \right]   < 0\,.
\label{eq:normality} 
\end{equation}
In words this says that $k \cot \delta_0(k)$ must cross the $-|k|$ line from below as $k^2$ becomes more negative, equivalently that the slope of $- \vert k \vert$ should exceed that of $k \cot \delta_0(k)$ at the crossing. This guarantees that the residue of the corresponding pole in the physical scattering amplitude, $\mathcal M_2$, has the proper sign, as dictated by inserting the bound state part of the identity, $\mathbb I = \vert E_B \rangle \langle E_B \vert + \cdots$, into its definition. One corollary is that, if there are two bound states, 
$k \cot \delta_0(k)$ must diverge between them~\cite{Iritani:2017rlk}.
 We will see cases of this.
 
 \begin{figure}[H]
\centering 
\subfigure[ \label{fig:kcot2} $m a_0 = 2$]{\includegraphics[width=.70\textwidth]{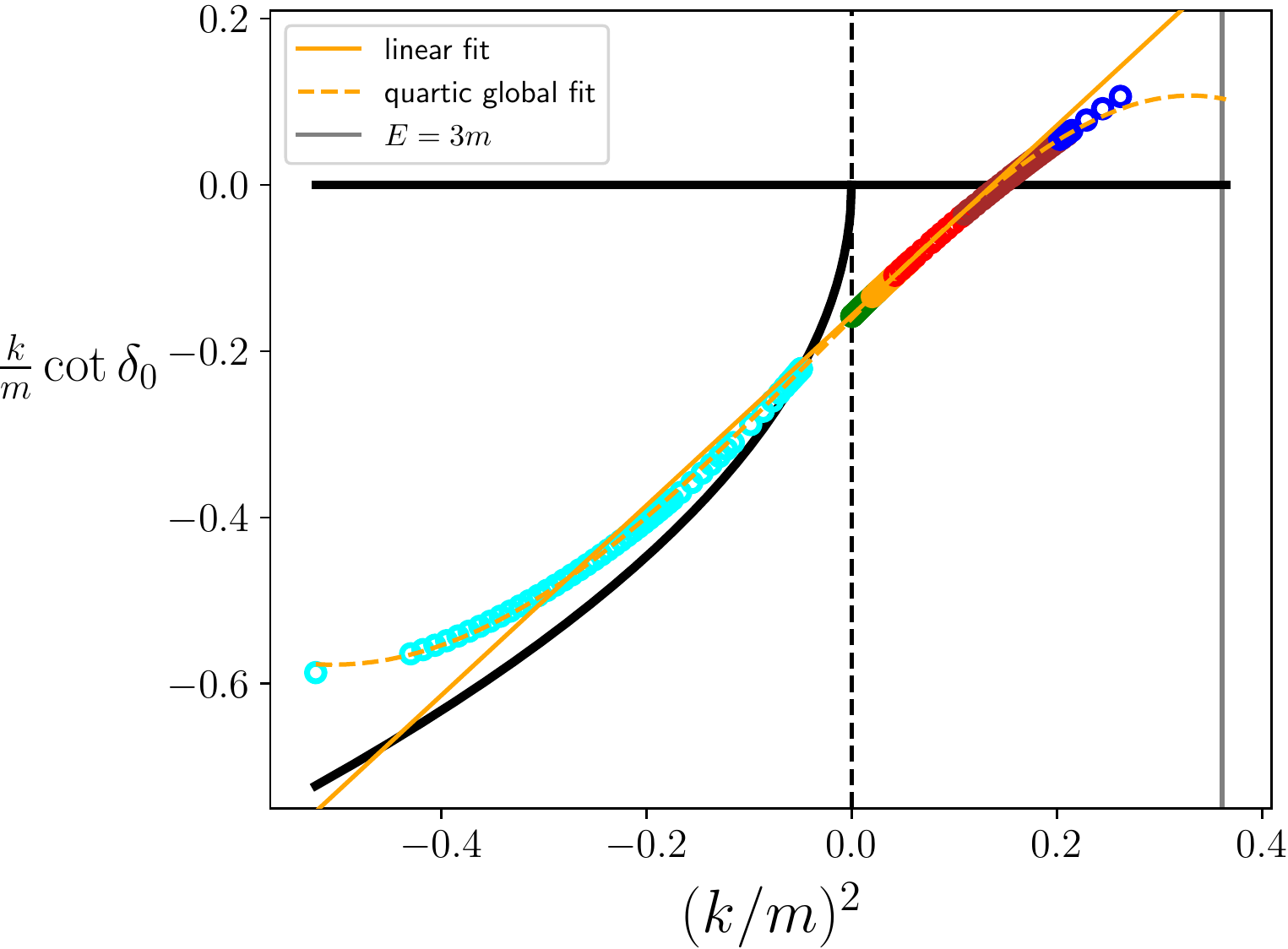}}
\hfill
\subfigure[ \label{fig:kcot6} $m a_0 = 6$]{\includegraphics[width=.49\textwidth]{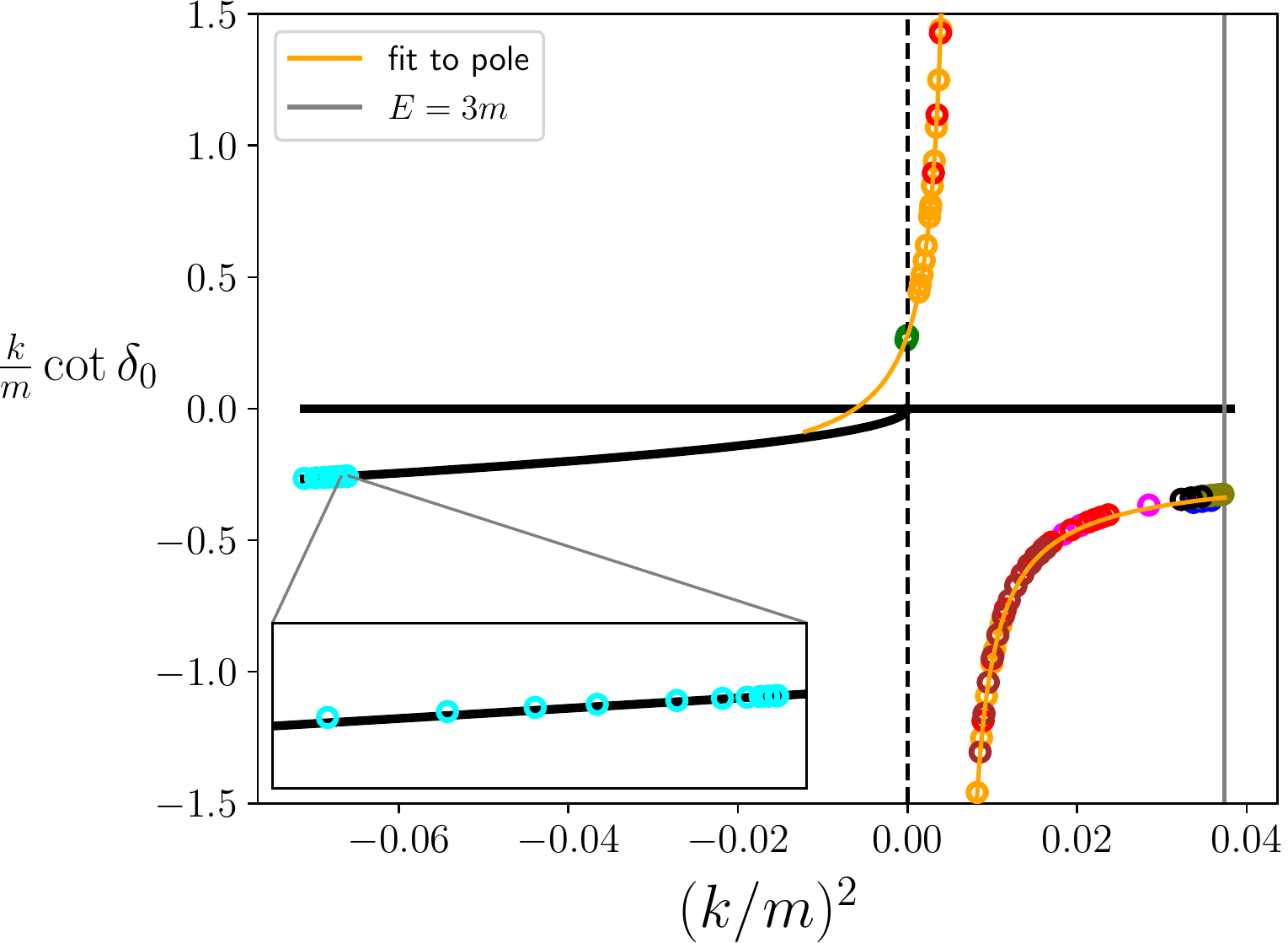}}
\hfill
\subfigure[ \label{fig:kcot16} $m a_0 = 16$]{\includegraphics[width=.49\textwidth]{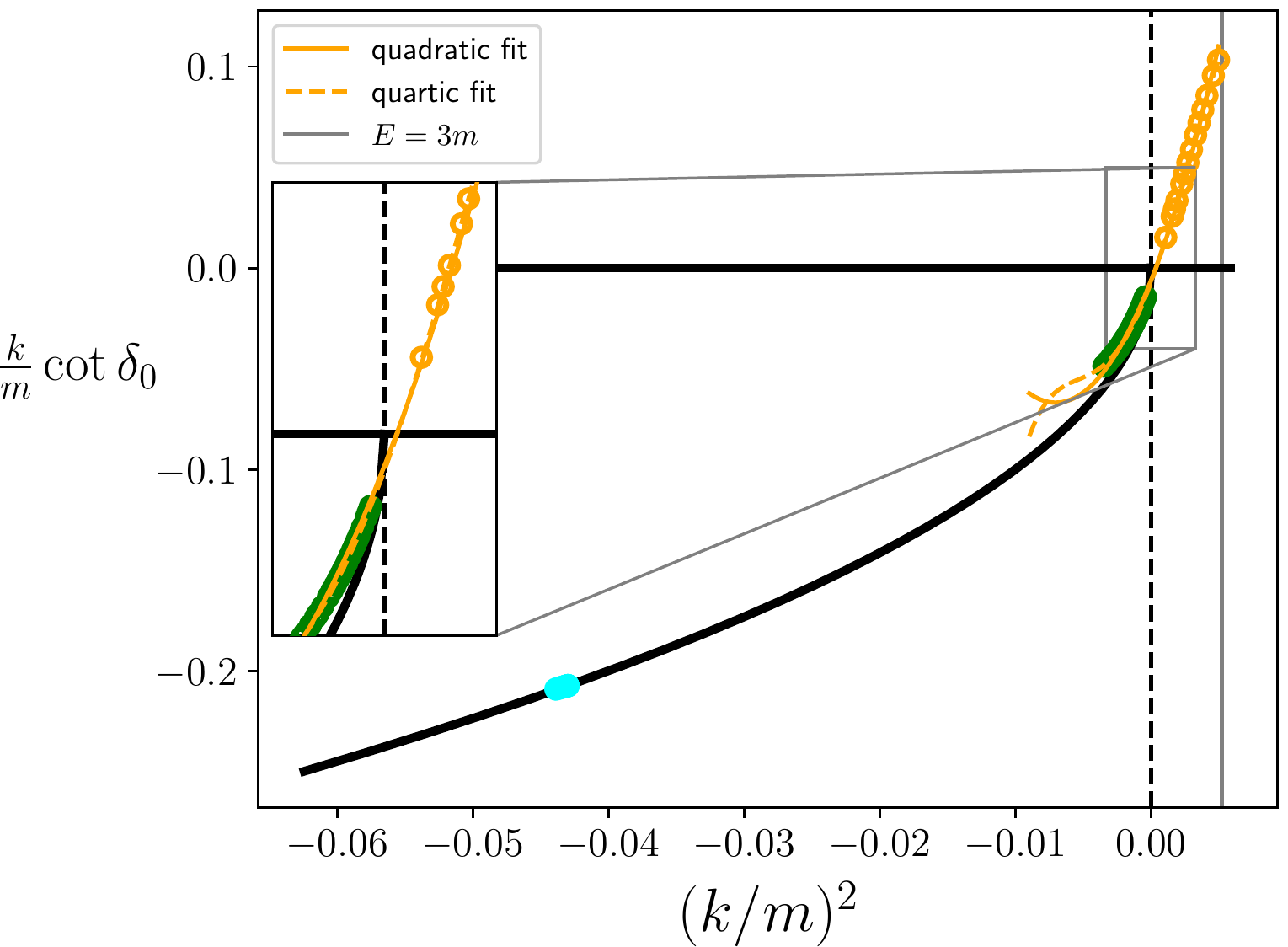}}
\caption{\label{fig:kcot} $k \cot \delta_0(k)$ for particle-dimer scattering
as a function of the relative center-of-mass momentum $k$, as determined from the three-particle
finite-volume spectrum in the $2+1$ regime using Eq.~(\ref{eq:QC2}).
We set $\Kiso=0$ throughout, and consider three different choices of $m a_0$.
Different colors represent data points coming from different states in the spectrum,
 and there is considerable overlap of points for $k^2 > 0$.
The solid vertical line near the right-hand edge of each plot shows the value of $(k/m)^2$ corresponding
to the three-particle threshold, i.e. $E=3m$. The curved solid line for negative $k^2$ shows $-|k| = - \sqrt{- k^2}$.
A bound state is present whenever $k \cot \delta_0(k)$ passes through this line.
The fits and other aspects of the plots are described in the text.}
\end{figure}
  
\medskip
We begin with the case of $m a_0 =2$, with a relativistic dimer lying
well below the three particle threshold, as discussed in Sec.~\ref{sec:largea0}.
The results for $k \cot \delta_0(k)$ obtained using Eq.~(\ref{eq:QC2}), are shown in Fig.~\ref{fig:kcot2}. 
The color coding matches that of the levels in Fig.~\ref{fig:am2largeL}:
the cyan points lying at $k^2 < 0$ come from the trimer state, while the other points come from
the levels lying above the $2+1$ threshold at $E/m=1+\sqrt{3}$.
We use volumes $mL=4-50$ for the lowest state, and $mL=20-70$ for the rest, so that all the states lie below $E=3m$ and satisfy $L/a_0\gg 1$.
We observe several important features:
\emph{(i)} The results for $k \cot \delta_0(k)$ from different levels overlap,
and are consistent.\footnote{%
We do not expect perfect consistency, since there are exponentially suppressed corrections
to the quantization condition that are not included. Indeed, if we zoom in on the plots, we
find that the overlap is not perfect.}
\emph{(ii)} The results from the trimer and the $2+1$ states 
 can be well described by a quartic order ERE curve, 
shown in the figure by the dashed line. 
We also show a linear fit (i.e. including $b_0$ and $r$) to the orange points to give an idea
of the rate of convergence of the ERE.
\emph{(iii)} The bound state that occurs where the ERE crosses $-|k|$ line is the trimer seen
in Fig. \ref{fig:spectra} above. The crossing occurs in the physical direction required by Eq.~(\ref{eq:normality}).
\emph{(iv)} The inclusion of higher order terms in the ERE is essential to describe our results.
Thus, although the underlying two-particle interaction is described exactly by the leading order
term in its ERE (by construction), the resulting dimer-particle ERE shows more structure.
It is the result of solving the field-theoretic problem of particle scattering from a relativistic bound state.
In particular, the value of the dimer-particle scattering length $m b_0\approx 6$ is not close to
that for the underlying particles, $ma_0 =2$.
\emph{(v)} Finally, we note that, were we to truncate the ERE as shown by the solid line,  then there
would be a second bound-state crossing at $(k/m)^2 \approx -0.4$, but the direction of
crossing the $-|k/m|$ curve would be unphysical. This is avoided by the results themselves and
shows again the necessity of higher terms in the ERE.

We deduce from this first example that the energy levels for $E < 3m$ are well described
by a dimer $+$ particle effective theory, and that the trimer at this value of $m a_0$ should be
understood as a dimer-particle bound state. A somewhat similar, nearly-physical situation is $\pi \sigma$ or $\pi \rho$ scattering for quark masses at which  the $\sigma$ or $\rho$ is stable.
The latter case has been  as studied in Ref.~\cite{Woss:2018irj}. The current quantization condition cannot, however, directly address either of these cases, due to the restriction to identical particles.  

The second example we study is that with $m a_0=6$. The results for this case are shown in Fig. \ref{fig:kcot6}. 
The corresponding dimer lies much closer to threshold, $M_d=1.97 m$, as can be seen from
the smaller range of $(k/m)^2$ available below the three-particle threshold.
Nevertheless, by going to large volumes, with $mL=60-170$, we are able to determine $k \cot \delta_0(k)$ from seven levels and fill out the curve for $k^2 > 0$. If we use smaller values of $L$, then the results begin to depart from the universal curve, due to large finite-volume effects on the dimer itself.\footnote{%
These can be partially removed by using the quantization condition with $k$ determined using
the volume-dependent dimer mass, $M_d(L)$, but we have not pursued this approach as we are able
to work directly for values of $L$ for which $M_d(L)-M_d$ is extremely small.}

The result for $k \cot \delta_0(k)$ is significantly changed from that with $m a_0=2$.
For one thing, the dimer-particle scattering length has changed sign to $m b_0\approx -3.6$,
corresponding to a moderate attraction and no bound state (trimer) near threshold.
For another, there is a pole that limits the range of applicability of the standard ERE to a tiny region around threshold. The presence of such a pole  indicates only that the phase shift is passing through
$0\ {\rm mod } \ \pi$; it is similar to that seen in physical nucleon-deuteron scattering, as discussed in Sec.~\ref{sec:nD} below. 
We find that using the modified ERE of Eq.~(\ref{eq:EREpole}), we obtain
 an excellent description of the results around and above threshold. 
 This is shown in the figure by the orange curve, which is a fit only to the orange points (the third
 energy-level in the spectrum), so as to emphasize the consistency with the results from the other
 levels.
 The exception to this consistency are the results from the trimer, shown again in cyan. Although
 difficult to see from the figure, these points do pass through the $-|k/m|$ line in the physical
 direction. In order to avoid an unphysical crossing, these points can be connected to the orange
 curve only if there is an intervening pole. 
 Thus, while the trimer will appear as a pole in the dimer-particle scattering amplitude, 
 the behavior of the phase shift is not given by a simple function, unlike in the previous case.

The last example studied in detail is that for $ma_0=16$, 
with results shown in Fig.~\ref{fig:kcot16}. 
The dimer is now very shallow, $M_d=1.996$, so the energy regime described
by dimer $+$ particle states is much reduced. Because of this, and the need to have $L\gg a_0$,
we have results for only one level above threshold (which itself requires $mL>80$).
As we saw in Sec.~\ref{sec:largea0}, there are now two trimers, 
one shallow (the green points) and the other deeper (those in cyan).
A quadratic fit to the orange points, shown by the solid line,
correctly  determines the bound state energy,  with $mb_0\approx 100$.
But, as for $ma_0=6$, the ERE has a small radius of convergence and cannot describe all the
results. Our interpretation is that the form of $k \cot \delta_0(k)$ is qualitatively similar to 
that for $m a_0=6$, but with the pole moved to the right so that $b_0$ changes sign.\footnote{%
Comparing Figs.~\ref{fig:kcot6} and \ref{fig:kcot16} allows us to understand in more detail
the issue of the missing level discussed at the end of the previous subsection. The quantization
condition (\ref{eq:QC2}) is satisfied whenever $k \cot \delta_0(k)$ equals the (appropriately rescaled)
L\"uscher zeta function. Far below threshold, the right-hand side of the quantization condition
asymptotes to the line $-|k|$, while it approaches $+\infty$ as $k^2\to 0$. Thus, at fixed $mL$,
as one lowers the $k \cot \delta_0(k)$ curve (moving, say, from the shape seen in Fig.~\ref{fig:kcot6}
to that of Fig.~\ref{fig:kcot16}) the lowest solution to the quantization condition
will vary continuously from a $2+1$ scattering
state (with $b_0 < 0$) to a shallow bound state (with $b_0> 0$). No additional finite-volume
state appears as $b_0$ crosses zero.
}
Again the deeply bound trimer cannot be viewed as a simple dimer-particle bound state.

To conclude this subsection, we show in Fig.~\ref{fig:b0}
the dependence of the particle-dimer scattering length, $b_0$,
on the underlying two-particle scattering length, $a_0$.
This allows us to understand the results for $m a_0=2$, $6$, and $16$ in a broader context.
For $m a_0 < 1.4$ (the left-most two points in the plot) there is only a moderate attraction
between particle and dimer (corresponding to $b_0 < 0$) and no trimer.
As $m a_0$ increases, $b_0$ has a pole and changes sign. For $m b_0 \gg 1$, we
expect there to be a shallow trimer that can be interpreted as a $2+1$ bound state, 
and the results for $m a_0=2$ show an example of this.
As $m a_0$ increases further, $m b_0$ decreases, and the trimer becomes increasingly bound,
as exemplified by the results at $m a_0=6$.
Continuing further, there is another pole in $m b_0$, above which a second shallow bound state
appears, as we have seen at $m a_0=16$.

For large $m a_0$, the dimer is nonrelativistic, and thus we expect that NREFT can be used
to study the $2+1$ system analytically.
We describe in Appendix \ref{app:NREFT} how this can be done  in the isotropic low-energy approximation, 
with only one free parameter corresponding to the three-particle contact interaction, or, equivalently,
the cutoff $\Lambda$.
The solid curve in Fig.~\ref{fig:b0} shows the result after tuning the cutoff so that the curve
matches the results from the quantization condition at large $ma_0$.
It describes our results well down to $m a_0 \approx 2$, where we expect relativistic effects
to become important.
This comparison provides another crosscheck of our formalism, while also showing where
relativistic effects are important.

The properties of the trimers are well studied and understood in NREFT. In particular, as we approach
unitarity ($m a_0\to \infty$) a sequence of Efimov bound states will appear. Thus we known that the
curve in Fig.~\ref{fig:b0} will have an infinite sequence of poles, separated asymptotically by a 
factor of $\approx 22$ in $ma_0$ \cite{Efimov:1970zz,Bedaque:1999ve}.
Turning this around, we can interpret the appearance of the second trimer seen at $m a_0=16$
as the second state in the Efimov sequence.
 We note however that the separation of the first and second trimer in $ma_0$ is smaller than the NREFT prediction due to relativistic effects---with the ratio given by $\approx 9$, rather than $\approx 22$.

\begin{figure}[H]
\begin{center}
\includegraphics[width=.60\textwidth]{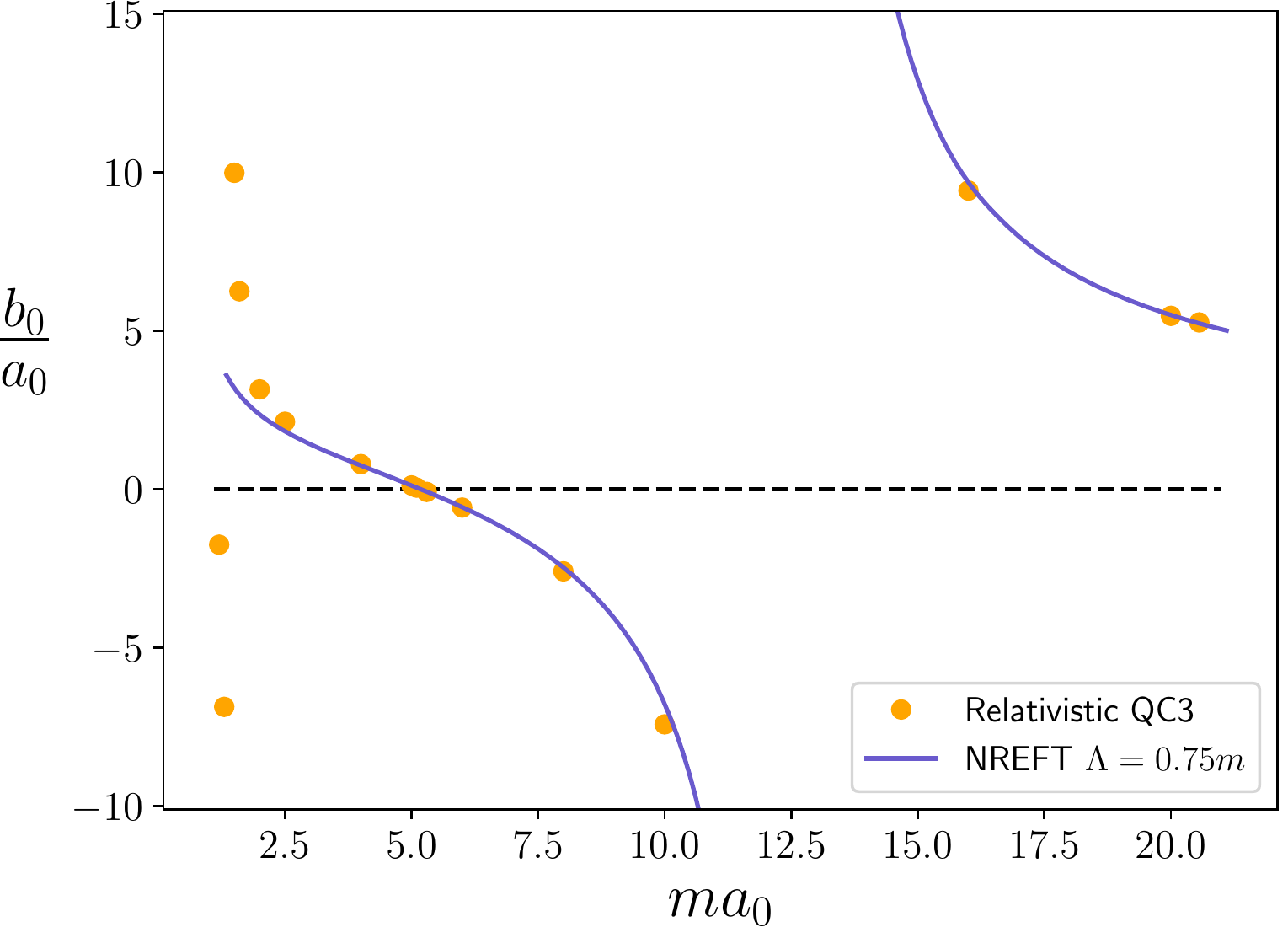}
\caption{
Ratio of particle-dimer scattering length, $b_0$, to the fundamental scattering length, $a_0$,
as a function of $m a_0$. Orange points give the results obtained from the analysis 
described in this subsection applied to the output of the three-particle quantization condition (QC3).
The solid (blue) line is the result from (infinite-volume) NREFT, discussed in more detail
in Appendix \ref{app:NREFT}. The rapid variation of points on the left-side of the plot correspond to a narrow pole in the relation between $b_0$ and $a_0$  that only arises in the relativistic theory.
}
\label{fig:b0}
\end{center}
\end{figure}

\subsection{Tuning toward a physical system: a model of neutron-deuterium scattering}
\label{sec:nD}
So far we have only considered the effects of two-particle interactions on the finite-volume spectrum. 
In this section we go beyond this restriction by studying how
non-vanishing values of $\Kiso$ affect 
the finite-volume spectrum in the isotropic approximation, and in particular
 how these impact the particle-dimer phase shift. To explore this, 
we consider a toy model that mimics three-nucleon interactions.
Specifically, we assume isospin symmetry,
so that proton-neutron and neutron-neutron interactions are identical, 
and ignore spin-dependent interactions.
In this way, we arrive at a system for which the current form of the three-body quantization condition
is applicable. 
We then tune the parameters $a_0$ and $\Kiso$ to match the physical system as closely as possible.
Since this requires nonvanishing $\Kiso$, we must use the modificed $\PV'$ pole prescription with
nonzero $I^s_\PV$. 
Following the discussion  of Sec.~\ref{sec:bounds}, we set $I^s_\PV=-1$.

\begin{figure}[h!]
\begin{center}
\includegraphics[width=.7\textwidth]{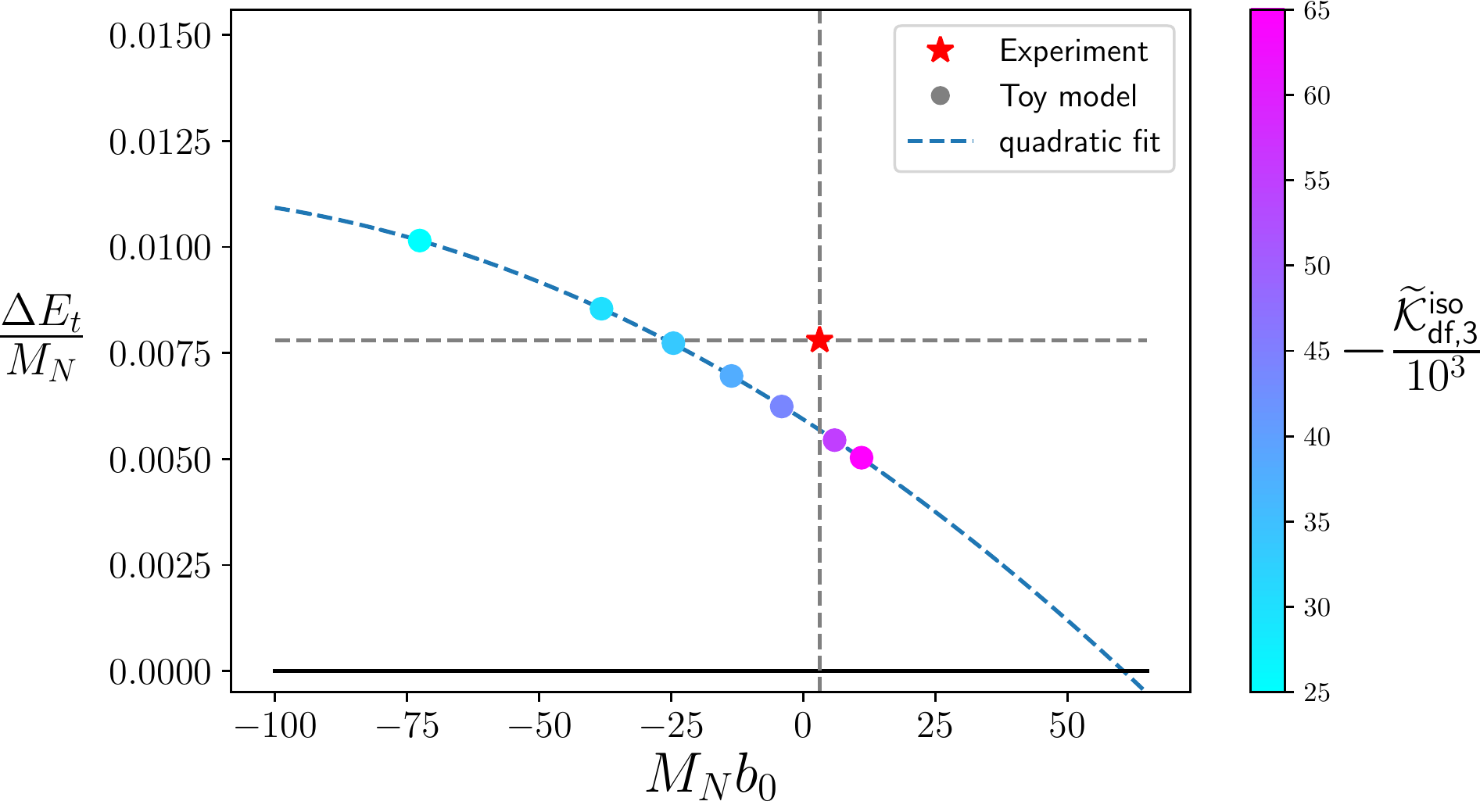}
\caption{Correlation between the triton (trimer) binding energy and the particle-dimer (nucleon-deuteron) scattering length. This is usually referred to as the Phillips line. The value of $\widetilde{\mathcal{K}}_{\text{df,3}}^{\text{iso}} = M_N^2 \Kiso$ used for each point can be determined from the color gradient at right. For further discussion see the text.}
\label{fig:phillips}
\end{center}
\end{figure}

{
We choose the value of the scattering length to reproduce the physical deuterium to nucleon mass ratio,
\begin{equation}
\frac{M_d}{M_N} \simeq 1.99763 \ \Rightarrow \ M_N a_0 = 20.56,
\end{equation}
where we set $m=M_N$, with $M_N$ the average of the proton and neutron masses.
As we can see from Fig.~\ref{fig:b0}, with $\Kiso=0$ this value of $a_0$ leads to two trimers.
To obtain a single trimer with mass close to that of the triton, $M_t$, it turns out that we
need large, negative values of $\Kiso$.
Figure~\ref{fig:phillips} shows the resulting tritium binding energy, $\Delta E_t=3 M_N-M_t$,
using a continuous color gradient to identify the value of $\Kiso$.
For each choice of this parameter, we also determine the particle-dimer scattering length using the
methods of the previous subsection. This allows us to plot $\Delta E_t$ vs. $b_0$ (in dimensionless
units), as shown in the figure.
These two quantities have been observed to be highly correlated in different potential models, following an approximate linear behavior known as the Phillips line \cite{\PhillipsOrig}.\footnote{%
See Fig.~13 of Ref. \cite{Hammer:2019poc} and surrounding discussion.
 The fact that potential models lead to results in the Phillips plot that lie in an  almost one-dimensional
 subspace was subsequently understood as being due to the fact that only one three-particle
 parameter is necessary at leading order in NREFT \cite{\PhillipsBHK}.
 The same explanation holds within our toy model, with its single three-particle parameter.}
We also include an experimental point, obtained with the physical values of $M_t$
and the neutron-deuteron scattering in the doublet (spin $1/2$) channel,
the latter taken from Ref.~\cite{Dilg:1971gqb}:
\begin{equation}
 \frac{M_t}{M_N} \simeq 2.9922 \,,  \ \ \ M_N b_0  \simeq 3.1.
\end{equation}
We observe that, even though our toy model cannot reproduce both experimental values simultaneously,
 the nearly linear dependence is qualitatively similar to the Phillips line shown in Ref.~\cite{Hammer:2019poc}. 
 It suggests that a sizeable three-particle interaction term is needed to understand the triton.

}

\begin{figure}[h!]
\begin{center}
\includegraphics[width=.60\textwidth]{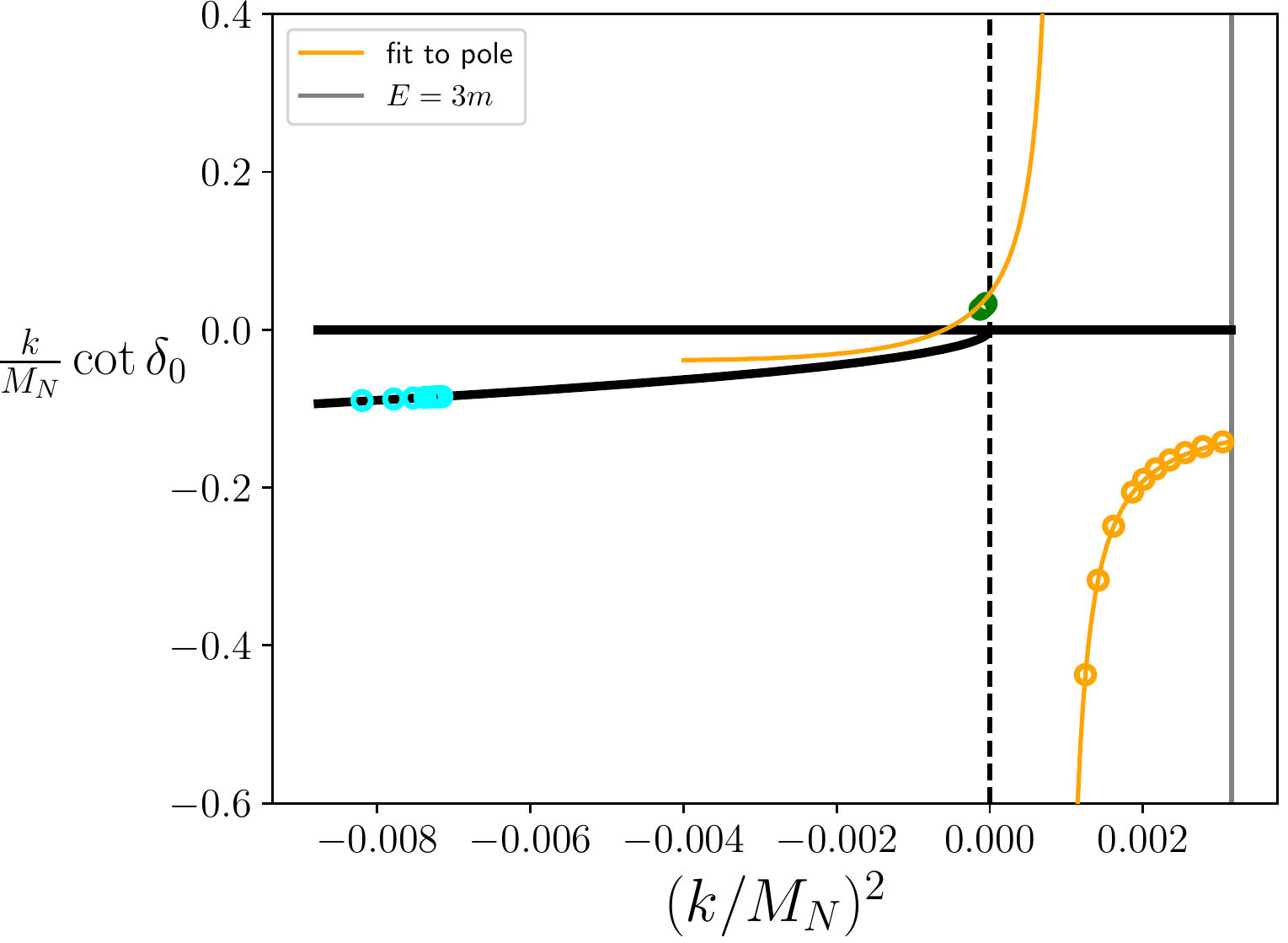}
\caption{$s$-wave phase shift as a function of the center-of-mass frame 
momentum for the simplified toy model describing 
nucleon-deuteron scattering. Parameters are $M_Na_0=20.56$ and
$M_N^2\Kiso=-33500$.  Notation as in Fig.~\ref{fig:kcot}. The fit is discussed in the text.}
\label{fig:NDscattering}
\end{center}
\end{figure}

We now study the phase shift of our toy nucleon-deuteron system in more detail. 
We choose to fix the tritium mass to its physical value, 
which will not reproduce the experimental scattering length, as already discussed. 
With $M_N^2\Kiso=-33500$ we find $ M_t = 2.99227 M_N$
which is close enough for the purposes of this work. In Fig.~\ref{fig:NDscattering} we show the resulting nucleon-deuteron phase shift, obtained using the method of the previous subsection. 
It is instructive to compare this plot to those shown in Fig.~\ref{fig:kcot}, which are obtained with $\Kiso=0$. 
Qualitatively, the present results are most similar to those at $m a_0=6$, Fig.~\ref{fig:kcot6},
despite the present value of $m a_0=20.56$ lying closer to $m a_0=16$.
This can be understood as follows:
as one increases $a_0$ while keeping $\Kiso=0$, 
the pole moves to higher energies and a second bound state emerges. 
Turning on a negative (and thus repulsive) $\Kiso$, the pole is moved to lower values of $k^2$ 
and the shallow bound state smoothly turns into a scattering state. 
We thus see that the differences from the $m a_0=16$ results of Fig.~\ref{fig:kcot16}
are mainly due to the three-particle interaction.

Fitting the orange points in Fig.~\ref{fig:NDscattering} to the modified ERE of Eq.~(\ref{eq:EREpole}), 
we find that it provides an excellent description, 
including the green points from the level close to threshold.
It is interesting to compare this to experimental results for $N-D$ scattering.
There is indeed evidence of a pole $k \cot \delta_0(k)$ close to threshold in both $n-D$ and $p-D$
scattering,  although in the former it lies below threshold, while in the latter its position is not 
settled~\cite{Babenko2008,Black,rusos}. In fact, in our model the position of the pole can be inferred from Fig. \ref{fig:phillips}: the point for which $m b_0=0$ is when the pole in $k \cot \delta_0(k)$ is at threshold. 
We speculate that we could
 further tune our model by adding an energy dependence to $\Kiso$ 
 in such a way that the pole shifts to lower energies, while keeping $M_t$ constant.
However, this is beyond the scope of our already simplified example.

In summary, a simplified model with two parameters 
is able to reproduce qualitative features of the nucleon-deuterium phase shift, 
such as the presence of only one bound state and a pole in $k \cot \delta_0(k)$. Furthermore, it suggests that a repulsive three-body force is necessary to explain the dynamics of the system. It is thus a good example of how one could use the quantization condition to solve the infinite-volume dynamics 
of a realistic three-particle system.  
{ Of course, in the present instance the dynamics is nonrelativistic,
and NREFT calculations are much more advanced and realistic than our toy model.
The advantage of our approach, however, is that it works also in the relativistic domain.}

We conclude this section with a comment. Current lattice simulations with physical quark masses have volumes satisfying $M_N L \lesssim 30$. For such volumes the finite-$L$ effects on the deuteron are significant and thus, to study nucleon-deuteron scattering using such lattices, one cannot employ the effective two-body description used above.
Instead, one will require the full form of the three-body quantization condition to analyze lattice results even in the region $E<3 M_N$.

\subsection{Three-particle spectrum with resonances}
\label{sec:BWres}

\begin{figure}[h!]
\begin{center}
\includegraphics[width=.60\textwidth]{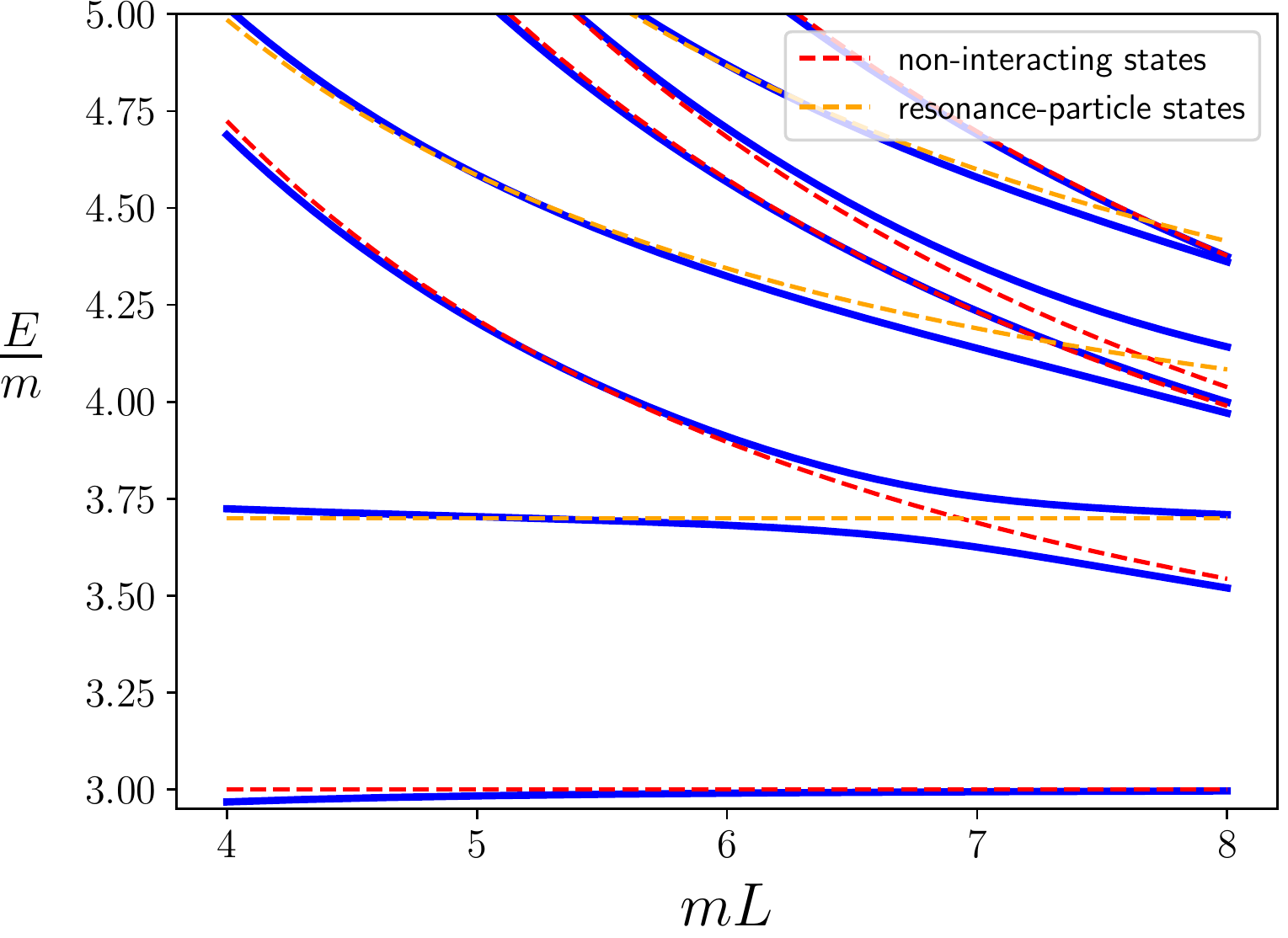}
\caption{Three-particle spectrum (solid blue lines) in the isotropic approximation in the presence of a two-particle resonance with $g=1$, $m_R = 2.7 m $ and $\Kdf=0$. 
Noninteracting three-particle states  (red dashed lines)
and noninteracting particle $+$ resonance states (orange dashed lines)
are shown for comparison. 
}
\label{fig:rho}
\end{center}
\end{figure}

{
The previous subsections focused on cases in which the two-particle channel had bound states. However, as explained in Section \ref{sec:BW}, the modified $\PV'$ prescription also allows the study of systems
in which two-particle subchannels are resonant. In this subsection we give an example of
the three-particle spectrum in such a situation.

Specifically, we use the parametrization of $\cK_2$ given in Eq.~(\ref{eq:BW}), with
$g=1$ and $m_R = 2.7 m$, and, for simplicity of implementation, set $\Kdf=0$.
The resulting spectrum is shown in Fig. \ref{fig:rho}. 
The first thing to notice is that there are additional
states compared to those expected for three almost free particles.
These extra states  can be interpreted as resonance $+$ particle states.
As in previous examples, there are avoided level crossings that occur when two states interchange
their interpretation---a clear case occurring for the second and third levels around $mL=7$.
Thus we find that, for a narrow resonance, a simple interpretation of the 
low-lying levels is possible. As in the two-particle case, however, for a broad resonance we expect that
a simple interpretation of levels will not be possible, and the
only way to interpret the spectrum is simply to use the full quantization condition and fit the parameters
contained in $\K_2$ and $\Kdf$.
}

\subsection{Including $d$-wave dimers}
\label{sec:numdwave}
\begin{figure}[H]
\centering 
\subfigure[ \label{fig:dwave1}$m a_0 = 0.1$, $m a_2 = 3.0$ and $\Kdf=0$. ]{\includegraphics[width=.49\textwidth]{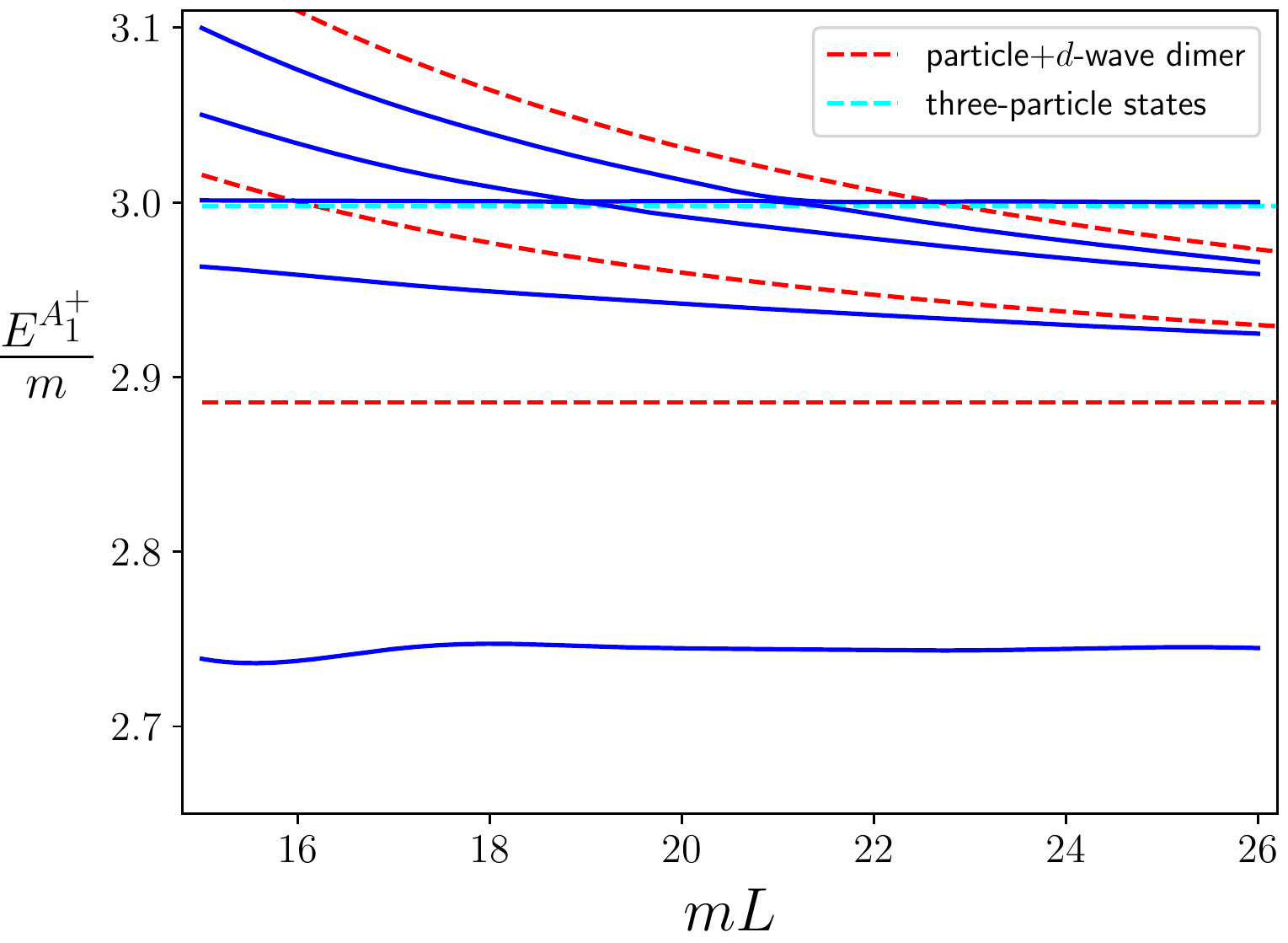}}
\hfill
\subfigure[ \label{fig:dwave2} $m a_0 = 2.0$, $m a_2 = 3.0$ and $\Kdf=0$. ]{\includegraphics[width=.49\textwidth]{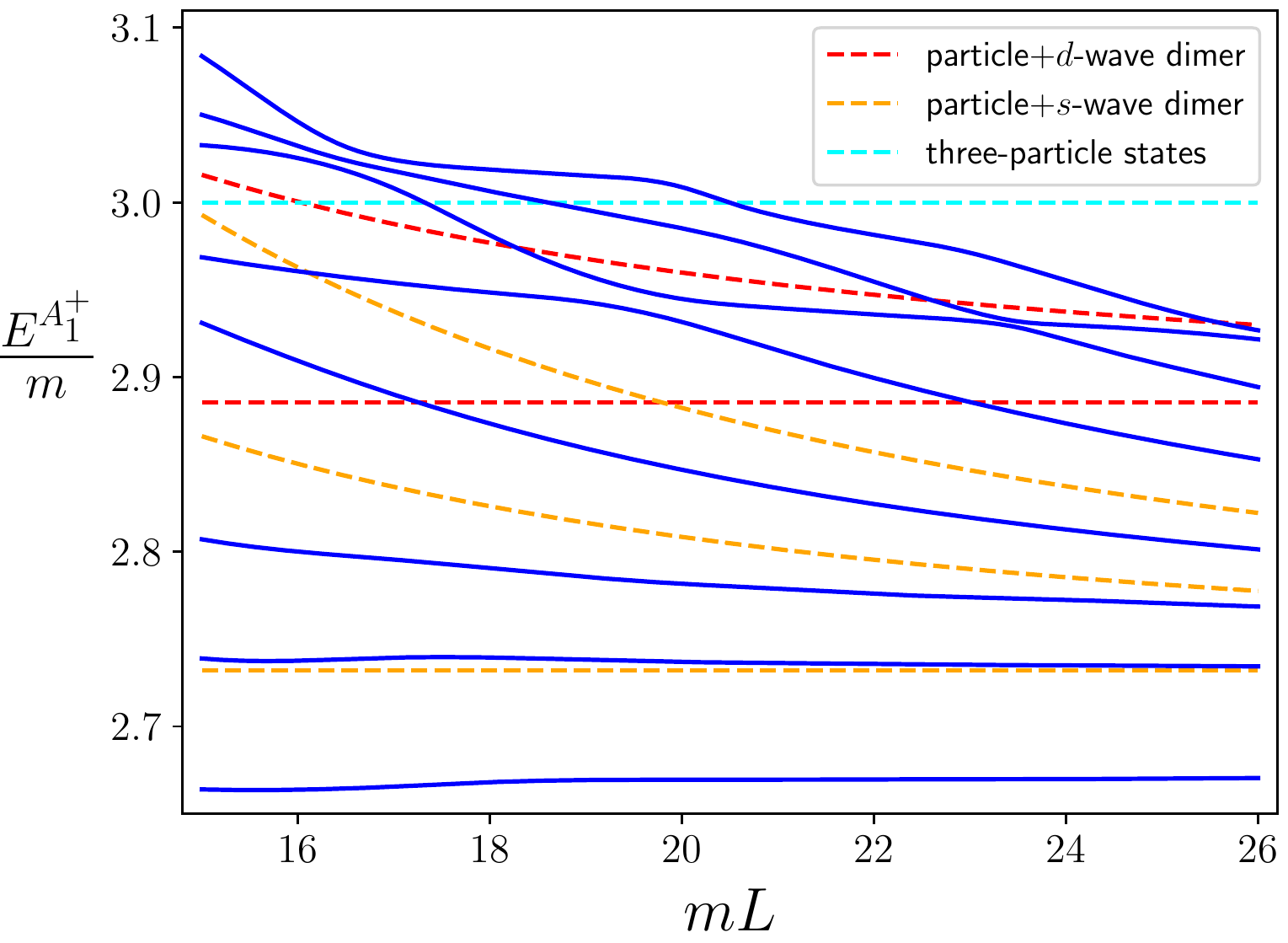}}
\hfill
\subfigure[ \label{fig:dwave3} $m a_0 = 2.3$, $m a_2 = 3.0$ and $\Kdf=0$. ]{\includegraphics[width=.49\textwidth]{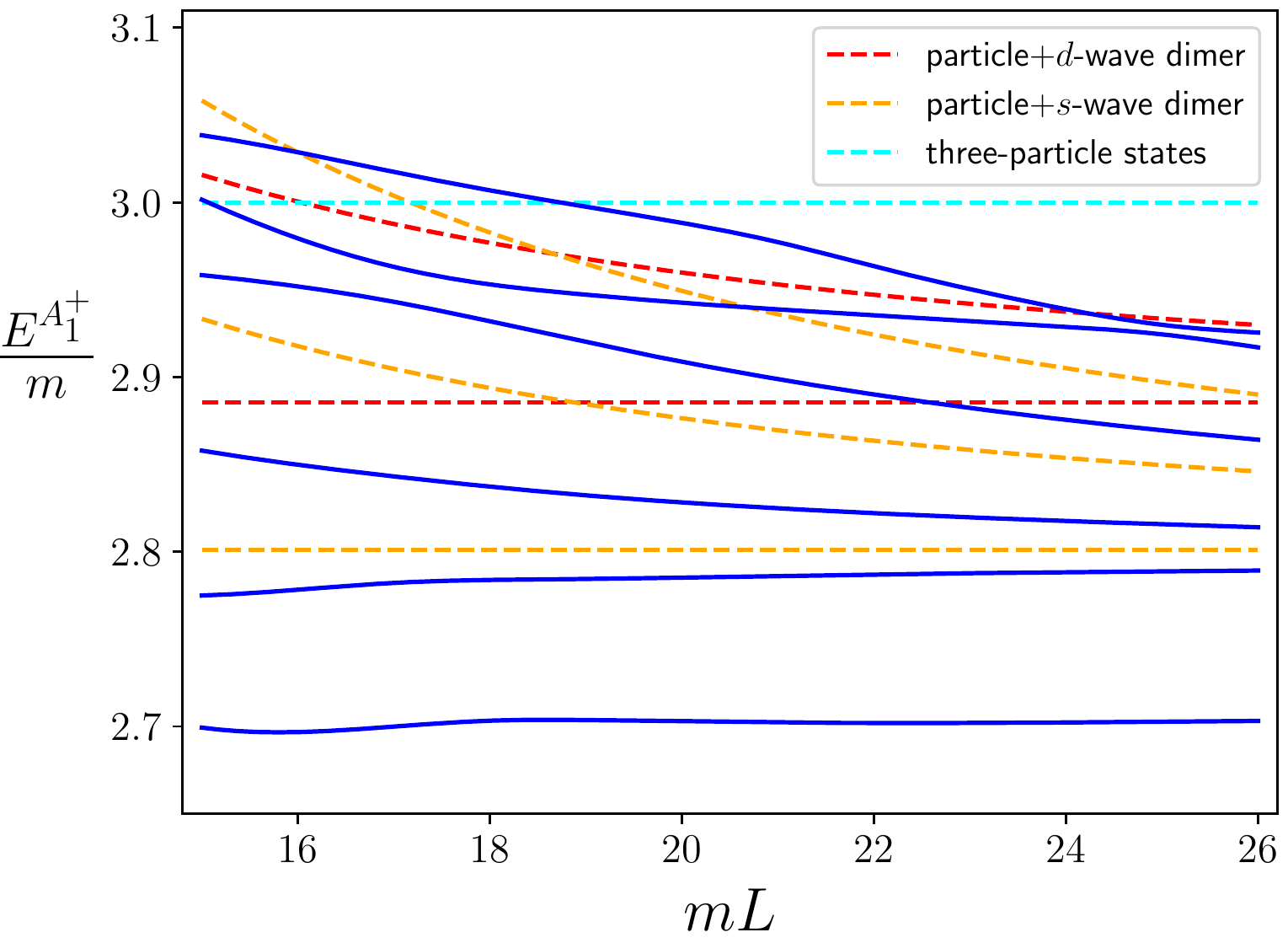}}
\caption{\label{fig:dwavetot} 
Finite-volume spectra (solid blue lines) of the three-particle systems in the $\mathbb A_1^+$ irrep with both
$s$- and $d$-wave two-particle interactions. Noninteracting three-particle levels,
as well as those involving particle $+$ dimer, are also shown. 
(In the upper right panel, the cyan dashed line at $E=3m$ has been shifted slightly downward to make it visible.)
See text for further discussion.}
\end{figure}
{
In our final numerical example we move beyond the isotropic approximation and include
both $s$- and $d$-wave two-particle channels. 
This setup has been studied previously in Ref.~\cite{\dwave}, but only for parameters such
that there are no dimers and no subchannel resonances.
Our aim is to investigate the spectrum
in a situation that is more akin to those that arise in nature, i.e.~with multiple two-particle channels
in which there are bound states.
As explained in Sec.~\ref{sec:boundd}, the $\PV'$ prescription allows us to study such systems
using the quantization condition of Ref.~\cite{\HSQCa}, 
by introducing a different $I_\PV^{(\ell)}$ for each partial wave.
Including $I_\PV^{(\ell)}$ in the implementation of Ref.~\cite{\dwave} is straightforward.
In fact, we will consider here only the choice $\Kdf=0$, for which we can use the original
implementation without change.

We consider only the simplest nontrivial extension of the examples in previous sections,
in which both $s$- and $d$-wave scattering are described by the lowest terms in their
respective threshold expansions, Eqs.~(\ref{eq:effrange}) and (\ref{eq:K2td}). 
Since we also set $\Kdf=0$, there are only two parameters: the scattering lengths $a_0$ and $a_2$.
We focus on values of the latter such that $m a_2 >1$,  
implying that there is an infinite-volume tensor bound state at
\begin{equation}
M^{	\ell=2}_d = 2m \sqrt{1 - 1/(m a_2)^2}\,.
\end{equation}
For the sake of brevity, we consider only three-particle states lying in the $\mathbb A_1^+$ irrep, although we stress that our implementation allows one to
study all available irreps, as shown in Ref.~\cite{\dwave}. 
It is important to keep in mind in the following that, due to the possibility of switching the spectator particle, the contributions of $s$- and $d$-wave subchannels are coupled, even in infinite volume.\footnote{%
For example, if the total angular momentum is $J=0$ (which is the dominant contribution in the
 $\mathbb A_1^+$ irrep in the case of $\vec P=0$), this can be produced both by an $s$-wave dimer with
 orbital angular-momentum $\bar \ell=0$ relative to the spectator, or by a $d$-wave dimer with $\bar \ell=2$.}

The first case we analyze is $m a_0 = 0.1$ and $m a_2 = 3.0$, leading to the spectrum
shown in Fig. \ref{fig:dwave1}.
For these parameters there is a $d$-wave dimer with $M^{\ell=2}_d  \approx 1.886 m$ but no
$s$-wave dimer.
Although such a situation may be unphysical (since in NRQM a potential with a $d$-wave bound state 
would also have at least one $s$-wave bound state), it is a simple starting point for studying the
finite-volume spectrum.
The spectrum shows a deeply bound trimer, similar to that observed in Ref.~\cite{\dwave} for
$m a_2 \approx -2$. In addition, there are several states that can be interpreted as
particle $+$ $d$-wave dimer scattering states, 
and which behave similarly to those in the pure $s$-wave case discussed earlier. 
We see also several strikingly-narrow avoided level crossings for $E\approx 3m$
(whose nature as avoided crossings can only be seen on a magnified version of the plot).
This narrowness is due to the very weak $s$-wave scattering length, so that, away from the crossings,
the level at $E\approx 3m$ is nearly a noninteracting state of three particles at rest.
We observe that the trimer energy has small oscillations, which are similar to those seen
in Ref.  \cite{\dwave}, and are likely indicative of unphysical effects arising  from the truncation
of the quantization condition or the enhancement of exponentially-suppressed effects.
These deserve further investigation, but this is beyond the scope of this paper.

A more physical situation is when there are both $s$- and $d$-wave dimers,
with the former being more deeply bound.
 With this in mind, we explore the effect of increasing $m a_0$ while holding $m a_2$ fixed, plotting the spectrum for $m a_0 =2$ and $m a_2 =3$  in Fig.~\ref{fig:dwave2}, 
 and for $m a_0 =2.3$ and $m a_2=3$ in Fig.~\ref{fig:dwave3}.
 We take these two different choices of $a_0$ in order to help clarify the interpretation of the spectrum.
 In Fig.~\ref{fig:dwave3} we clearly see two trimers. 
 We interpret the lower one as $s$-wave dominated (and thus similar to the trimers seen in earlier
 subsections) since it becomes more deeply bound for $m a_0=2$.
As for the upper trimer in Fig.~\ref{fig:dwave3}, we conjecture that it is primarily caused by the
$d$-wave attraction. 
This is based on the observation that it smoothly transforms into the $d$-wave
trimer of Fig.~\ref{fig:dwave1} as $m a_0$ is decreased. 

To make these characterizations rigorous it would be instructive to study the pole positions of these two trimers in the scattering amplitudes of the (scalar dimer + particle) $\leftrightarrow$ (tensor dimer + particle) coupled-channel system. In particular the set of two channels leads to the appearance of four Riemann sheets, conveniently labeled by the sign of the imaginary part of momentum carried by each element of the back-to-back particle-dimer pair. For example the second sheet is defined by $\text{Im} k_{M_{0} + m} < 0$ and $\text{Im} k_{M_{2} + m} > 0$ and poles on the lower half of this sheet are close to physical scattering energies and are often interpreted as bound states (or molecules) built from the constituents of the heavy channel---in this case a $M_d^{\ell=2}$+$m$ molecule. The interpretation follows from noting that, if the lighter channel were turned off, the pole would move to the real axis of the physical sheet and thus become a physical bound state pole. This behavior was observed for the $f_0$-like state in the scalar-isoscalar LQCD calculation presented in Ref.~\cite{Briceno:2017qmb}. Performing a coupled channel analysis here to extract the upper trimer pole position goes beyond the scope of this work, but would be an interesting future application of these results.

 The higher levels in Fig.~\ref{fig:dwavetot} appear to be predominantly particle $+$ $s$-wave dimer states, 
but there are some clear avoided crossings which we interpret as levels changing their nature
to particle $+$ $d$-wave dimer states. The situation becomes even more complicated for
$E> 3m$, where three-particle components become relevant.

 We conclude by noting that, while the examples considered here are not directly relevant to
 hadronic physics, they may be of relevance to the physics of cold atoms.

\section{Conclusions\label{sec:conclusions}}

In this work we have presented an extension of the formalism of Ref.~\cite{\HSQCa} 
that allows the study of three-body systems in the presence of  two-body resonances or bound states.
 This removes a major shortcoming of the original formalism,
 which had previously only been resolved by a more complicated approach requiring the
 introduction of a fictitious two-body channel for each resonance~\cite{\BHSK}.
 In addition, our extension may provide an alternative for the $2 \to 3$ scattering formalism derived in Ref.~\cite{\BHSQC}. 
  We provide here only an intuitive explanation of the new extension; a derivation will be presented in Ref.~\cite{inprog}, along with a discussion of the relation to the work of Refs.~\cite{\BHSQC,\BHSK}.
  We stress that, with this extension, the formalism for $s$-wave dimers with general two-particle
  interactions is now of similar complexity to implement as that obtained from the other approaches 
  (NREFT and FVU), while being the only one worked out explicitly for higher partial waves. 
 
The extended formalism  can be implemented numerically with only minor changes to the
methods developed for the original formalism in Refs.~\cite{\BHSnum,\dwave}.
This has allowed us to 
present several examples of the influence of two-particle bound states and
resonances on the finite-volume three-particle spectrum,
including a case in which both $s$- and $d$-wave interactions are included.

We have also presented several examples where the three-particle quantization condition can
be used to study infinite-volume physics, despite being originally formulated with finite-volume
applications in mind. The simplest example is the determination of the presence and binding
energies of trimers. We reproduce the expected Efimov-like trimers as the unitary limit of two-particle
scattering is approached, and can extend the results into the relativistic domain. 
We also find a complex
pattern of trimers induced by a combination of $s$- and $d$-wave two-particle attraction. 
What is most novel here, however, is that the quantization condition can be used to determine the
dimer-particle scattering amplitude for essentially all energies below the breakup threshold, 
reproducing expectations in the nonrelativistic regime and obtaining new results for 
relativistically-bound dimers. As an application, we study a toy model 
of the nucleon-deuteron-triton system, without spin or isospin,
which we find requires the use of  a nonvanishing value for the three-particle quasilocal interaction $\Kdf$.

With the extension presented here, we now have a relativistic formalism that is
straightforward to implement and can be used for
any system of identical scalar particles, with any (finite) number of two-particle partial waves.
In QCD, however, the only such system is  three pions with $I=3$, 
for which all subchannels, having $I=2$, are neither resonant nor have bound states.
The next step in the development of a generally applicable formalism is to include nonidentical
but degenerate scalars, which would allow the application to a general three-pion system
in the isosymmetric limit, and thus to the $\omega$, $a_1$, $a_2$ and other mesonic resonances.
This generalization is now one of our main priorities going forward.

One topic not directly addressed here is the use of the integral equations connecting $\Kdf$ to $\cM_3$.
We note, however, that the methods 
introduced in Ref.~\cite{\BHSnum} to solve these equations below or at threshold 
for the case without subchannel resonances or dimers should apply as well
in the presence of such resonances and dimers. They would allow us, for example, to study
the wavefunction of the triton in our toy model. 
We save such calculations until we can address a more physical example, rather than toy models.

Finally, an important issue that we have not addressed here is the presence of unphysical
solutions to the quantization condition for certain choices of parameters, 
as observed in Refs.~\cite{\BHSnum,\dwave}.
We are presently investigating whether these are removed by increasing the cutoff used
to truncate the sum over the spectator momentum. Resolving this is another major priority in order to provide a fully general tool for studying all possible three-particle systems.

\acknowledgments

We thank  Hans-Werner Hammer, Akaki Rusetsky and Jia-Jun Wu for useful discussions.
FRL acknowledges the support provided by the projects H2020-MSCA-ITN-2015/674896-ELUSIVES, H2020-MSCA-RISE-2015/690575-InvisiblesPlus, and FPA2017-85985-P. The work of FRL also received funding 
from the European Union Horizon 2020 research and innovation program 
under the Marie Sk{\l}odowska-Curie grant agreement No. 713673 and ``La Caixa'' Foundation  (ID 100010434, LCF/BQ/IN17/11620044). The work of TDB and SRS is supported in part by the United States Department of Energy (USDOE)
grant No. DE-SC0011637, in part by the DOE, while that of RAB is supported in part by
USDOE grant No. DE-AC05-06OR23177, 
under which Jefferson Science Associates, LLC, manages and operates Jefferson Lab.
RAB also acknowledges support from the USDOE Early Career award, contract de-sc0019229. 
SRS also acknowledges partial support from the Australian Research Council
through Grant No.~DP190102215, from the George Southgate Fellowship,
and from the International Research Unit of Advanced Future Studies at Kyoto University,
and thanks the Physics Department and the University of Adelaide and the Yukawa Institute of Theoretical Physics at Kyoto University for their hospitality while parts of this work were done.

\appendix
\section{$I_{PV}$-dependence of $\Kdf$ and the spectrum}
\label{app:IPV}

As explained in Sec.~\ref{sec:generalizing}, when we modify the PV prescription
according to Eqs.~(\ref{eq:Fellshift}) and (\ref{eq:Kellshift}),
the spectrum is formally unchanged for all volumes,
provided we make a suitable change to $\Kdf$.\footnote{%
As usual, this statement holds up to exponentially-suppressed corrections.}
In practice, however, this statement breaks down once we approximate the quantization
condition by truncating the sum over $\ell$.
We also note that the required change to $\Kdf$ is not known {\em a priori},
and must be determined numerically.
It is the purpose of this appendix to study, within the context of a concrete example,
the size of the required changes to $\Kdf$ and 
of the residual volume-dependence in the spectrum.

In our example we follow the numerical investigations of Sec.~\ref{sec:largea0} and
use the isotropic approximation with $\ell_{\rm max}=0$, so that the infinite-volume amplitudes
are parameterized by $a_0$ and $\Kiso$. We set $m a_0=0.1$ and consider two PV schemes:
the original one with $I_\PV^s=0$ and the modified one with $I_\PV^s=-1$.
The derivation of the quantization condition is valid for both choices (see Sec.~\ref{sec:bounds}).
We choose a large volume ($mL=30$) at which to match the spectrum by tuning the values
of $\Kiso(I_\PV^s; E)$, and then study the volume-dependence of the difference between spectral lines
at other volumes.

\begin{figure}[!htbp]
  \centering
  \includegraphics [width =0.7 \linewidth]{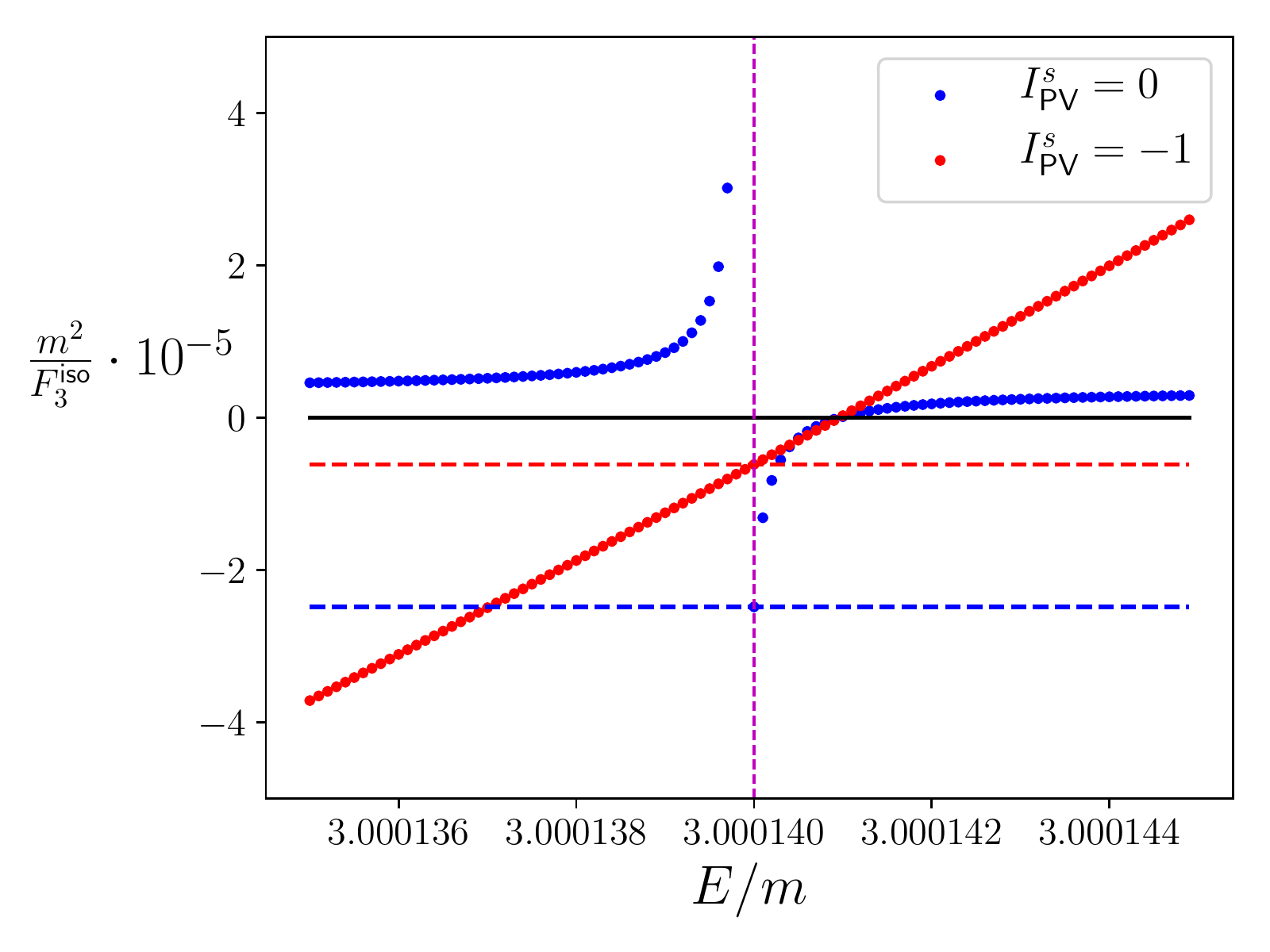}
  \caption{
  Determining the $I_\PV^s$ dependence of $\Kdf$ in the isotropic approximation using the
  ground-state energy, $E_0$, for $ma_0=0.1$ and $mL=30$.
 Blue and red points show $m^2/F_3^\iso$ for, $I_\PV^s=0$ and $-1$, respectively.
  The vertical purple dashed line shows our chosen tuning energy, while the horizontal dashed
  curves show the resulting values of $-\Kiso$. 
  }
  \label{fig:match_E0}
\end{figure} 

To show how this works we first consider the lowest energy level, $E_0(L)$.
In Fig.~\ref{fig:match_E0}, we show the energy dependence of the 
dimensionless quantity $m^2/F_3^\iso$ just above threshold for the two choices of $I_\PV^s$.
The quantization condition in the isotropic approximation is
\begin{align}
	 m^2\Kiso(  I_\PV^s; E) = - \frac{m^2}{F_3^\iso(I_\PV^s; E, L)}\,,
\end{align}
where we have made the scheme parameter explicit.
As discussed in Sec.~\ref{sec:generalizing}, we know that the solution to the quantization condition
is independent of $I_\PV^s$ if $\Kdf=0$.
This is seen in the figure by the fact that the two $m^2/F_3^\iso$ curves cross when they both vanish.
We are interested here, however, in cases where $\Kdf\ne 0$, and so  choose an energy
away from the crossing ($E_0=3.00014 m$, shown by the vertical line in the plot) 
and determine the values of $\Kiso$ so as to attain this energy. We find
\begin{equation}
\label{eq:Kiso_0}
\begin{split}
	m^2\Kiso(I_\PV^s=0;E_0)&=24.828\cdot10^4\,, 
	\\
	m^2\Kiso(I_\PV^s=-1;E_0)&=6.1365\cdot10^4\,. 
	\end{split}
\end{equation}

\begin{figure}[!htbp]
\centering 
\includegraphics[width=.70\textwidth]{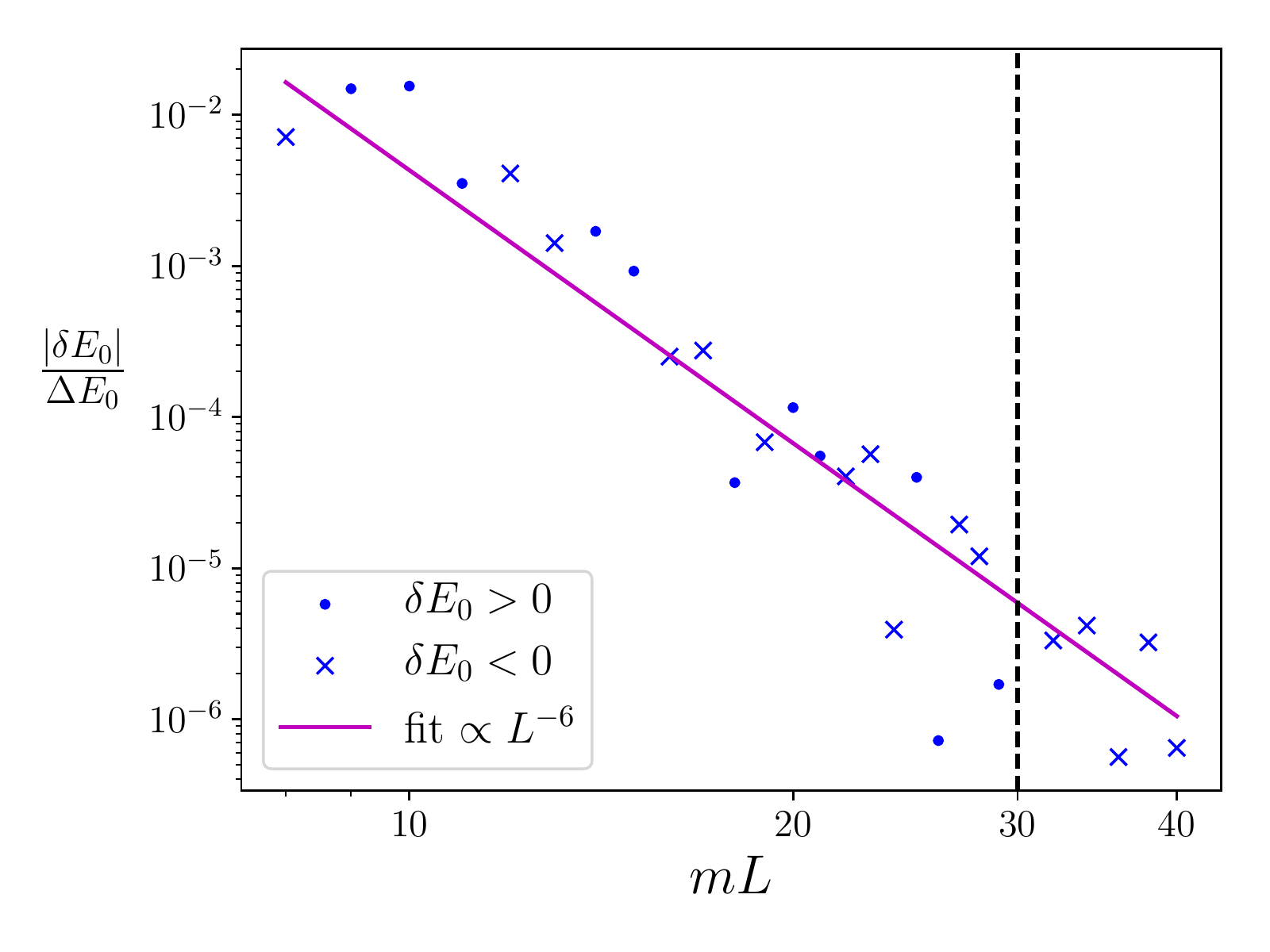}
 \caption{$I_\PV$-dependence of the ground-state energy $E_0(L)$, with $\delta E_0$ defined in 
 Eq.~\eqref{eq:delta_E0}.  
 We plot the ratio to $\Delta E_0 = E_0-3m$ in order to give a sense of the relative size of the shift.
 The vertical line at $mL=30$ indicates the point at which matching is performed, so that 
 $\delta E_0$ vanishes.
 \label{fig:ground}}
\end{figure}

We now determine the ground-state energies for other choices of $L$, keeping $\Kiso$ fixed
at the values given in Eq.~(\ref{eq:Kiso_0}).
We then evaluate the difference
\begin{align}
	\delta E_0(L) \equiv E_0(I_\PV^s=0;L) - E_0(I_\PV^s=-1;L)\,, \label{eq:delta_E0}
\end{align}
which is a measure of the residual $I_\PV^s$ dependence of the spectrum due to 
approximating the quantization condition.
The result is shown in Fig.~\ref{fig:ground}.
It oscillates about zero with an amplitude that decays rapidly with increasing $L$.

We can understand why the residual $I_\PV^s$ dependence is so small for the ground state
by using the threshold expansion developed in Ref.~\cite{\dwave}.
Close to threshold, the approximation of $\Kdf$ by an energy-dependent constant is valid
up to corrections of $\cO(\Delta)$, where
\begin{align}
	\Delta \equiv \frac{E^2-9m^2}{9m^2}\,.
\end{align}
Numerically, this is of $\cO(0.1\%)$ for the energies we are considering (choosing $mL\approx 10$).
Furthermore, what matters for $\delta E_0$ is the difference between the contributions of
the linear terms for the two choices of $I_\PV^s$, and the end result is the tiny effect
shown in Fig.~\ref{fig:ground}.
}

\begin{figure}[!htbp]
  \centering
  \includegraphics [width =0.7 \linewidth]{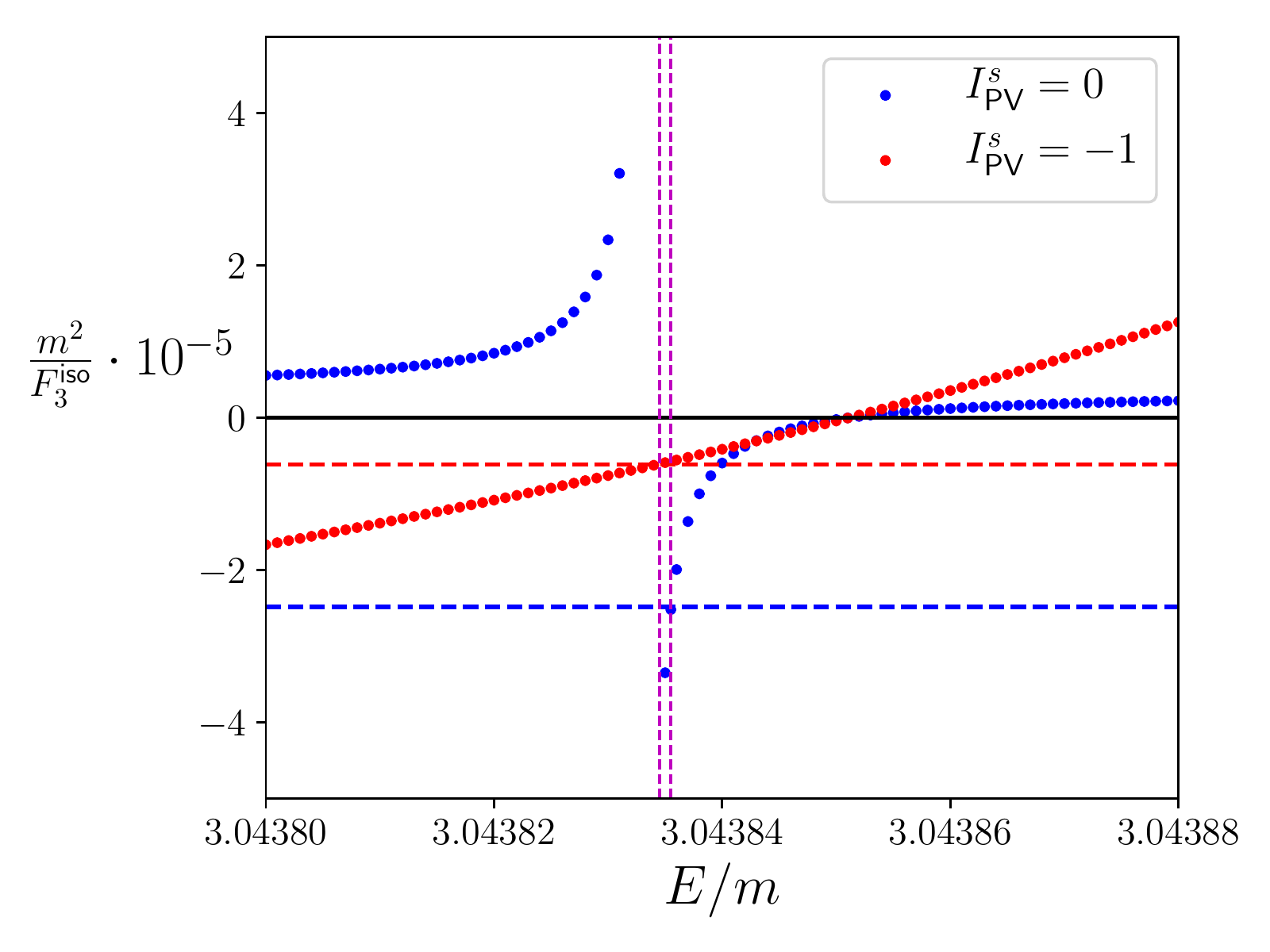}
  \caption{
  As for Fig.~\ref{fig:match_E0} but for the first excited state.
  The dashed magenta lines correspond to the values of $E_1(I_\PV=0;mL=30)$ and $E_1(I_\PV=-1;mL=30)$ obtained from the constant approximations $m^2\Kiso(I_\PV=0) = 2.4828\cdot10^5$ and $m^2\Kiso(I_\PV=-1) = 6.1365\cdot10^4$.
  The discrepancy between the two magenta lines is due to our neglect of the higher-order terms in the threshold expansion of $\Kdf$.
  \label{fig:match_E1}
  }
\end{figure}

We next extend our example to include the first excited state $E_1(L)$.
If we continue to assume energy-independent values of $\Kiso$,
then the first excited levels at $mL=30$ disagree by
\begin{align}
	\delta E_1(mL=30) \equiv E_1(I_\PV^s=0;mL=30) - E_1(I_\PV^s=-1;mL=30) \approx 10^{-6}m\,,
\end{align}
as shown in Fig.~\ref{fig:match_E1}.
Thus, {\em even at the same value of $L$}, the spectrum is dependent on $I_\PV^s$.
We interpret this as being due to our omission of the $\cO(\Delta)$ terms in the threshold expansion,
which are significantly larger for the first excited state than for the ground state.
As shown in Ref.~\cite{\dwave}, these still lead to an isotropic $\Kdf$, but now with linear energy
dependence,
\begin{align}
	 \Kdf(I_\PV^s;E) \approx \Kdfzero(I_\PV^s) + \Kdfone(I_\PV^s)\Delta\,.
	 \label{eq:Kdfone}
\end{align}
If we set $\Kdfone=0$ then we expect $\delta E_1(mL)\sim \Delta$.
In fact, we can make a more detailed estimate of $\delta E_1$ using the threshold expansion for
the excited state developed in Ref.~\cite{Pang:2019dfe}. 
The three-particle interaction enters the expression for $E_1$ first at $\cO(L^{-6})$. If this is mistuned
by $\Delta \sim 1/L^2$ then  we expect that $\delta E_1 \propto L^{-8}$.
This dependence is indeed what we find, as shown in Fig.~\ref{fig:deltaE1},

If we include the linear term in $\Kiso$, Eq.~(\ref{eq:Kdfone}), then truncation errors
are of $\cO(\Delta^2)$, so we expect this to perform considerably better. 
We set $\Kdfone(I_\PV^s)=0$, and then tune $\Kdfzero(I_\PV^s)$ and $\Kdfone(I_\PV^s)$
so as to set $\delta E_1(mL=30)=0$. 
We find that this requires $\Kdfone(I_\PV^s=-1)=4.123  \cdot 10^3$.\footnote{%
In practice we have left $\Kdfzero(I_\PV^s)$ unchanged and only tuned $\Kdfone$. This means that
$\delta E_0$ is slightly mistuned, but by an amount that is small on the scale of most of the
values shown in Fig.~\ref{fig:ground}.}
The resulting $\delta E_1$ is shown by the lower (red) points in Fig.~\ref{fig:deltaE1}.
Here we would expect the fall off to be as $L^{-10}$, since there are two extra powers of momentum.
We indeed find a fall off faster than $L^{-8}$, but with some oscillations that preclude detailed fitting. The detailed behavior and source of this dependence deserves a dedicated study that goes beyond the scope of this work. In such a future investigation we also intend to disentangle two possible sources for the residual $I_\PV^s$ dependence: {\emph{(i)}} The isotropic nature of $\Kiso$ is insufficient to keep the low-energy physics constant under variation of $I_\PV^s$; and {\emph{(ii)}} The quantization condition only holds up to neglected $e^{- m L}$ corrections and the exact form of these (and their imprint on the finite-volume solutions) varies with $I_\PV^s$.

\begin{figure}[!htbp]
  \centering
  \includegraphics [width =0.7 \linewidth]{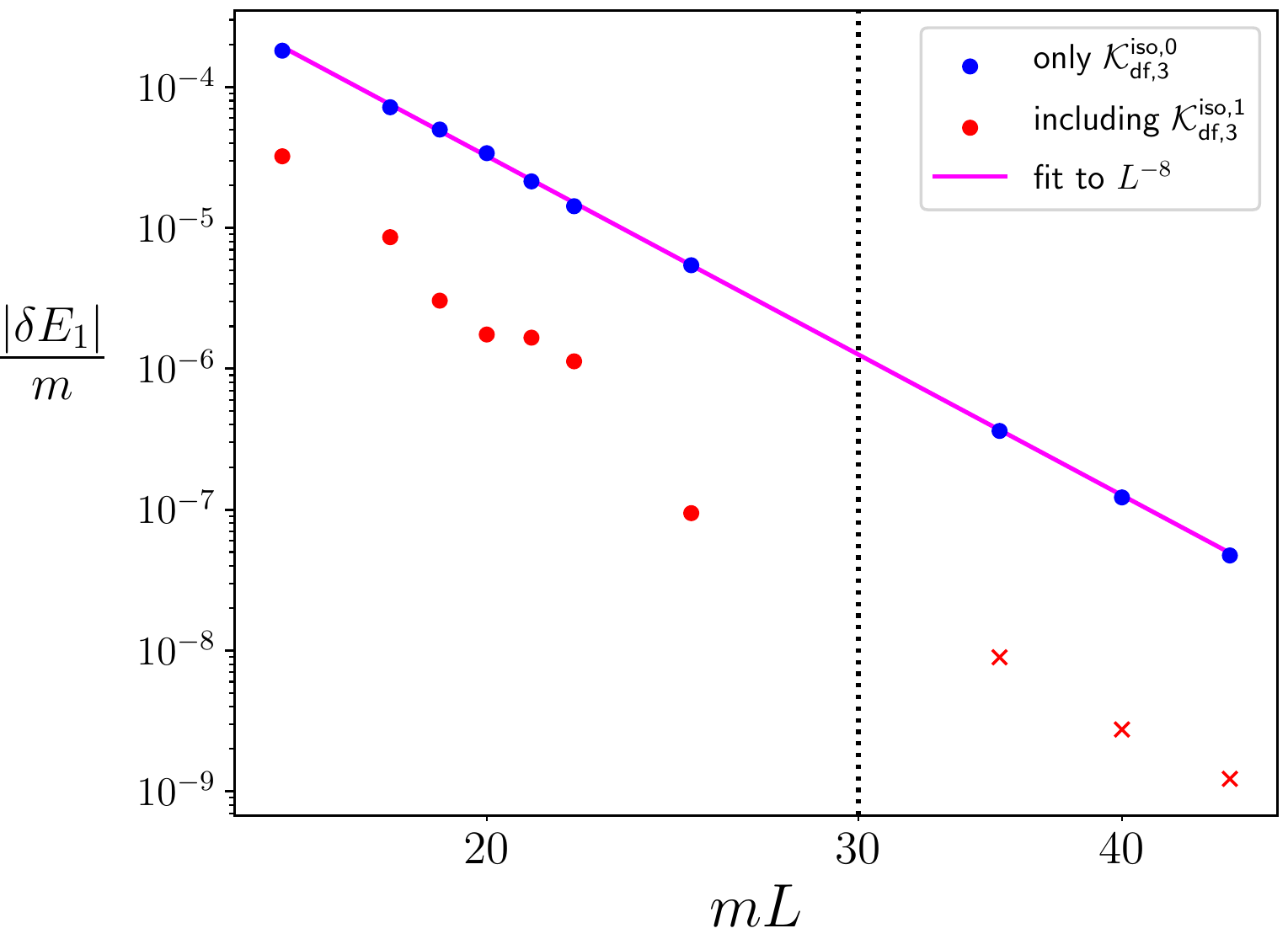}
  \caption{
 $|\delta E_1|$ as a function of $mL$ when tuning with only a constant $\Kiso$ (filled blue circles) or 
 with the linear dependence on $\Delta$ given in Eq.~(\ref{eq:Kdfone}) (filled red circles  and crosses).
 The solid (magenta) line shows a fit to the blue points assuming an $L^{-8}$-dependence.
 The vertical line indicates the value of $mL$ at which the tuning is done. Thus the corresponding
 red point vanishes at this value and is not shown. $\delta E_1$ is positive for the red circles and
 negative for the red crosses.
  \label{fig:deltaE1}
  }
\end{figure}

As a first step towards understanding \emph{(i)} we note that, in general, $\Kdf$ includes also non-isotropic terms that enter at $\cO(\Delta^2)$ (see Ref. \cite{\BRSd}). To summarize, in the main text we explained that modifying $I_\PV^s$ changes an isotropic $\Kdf$ into a nonisotropic one, and in this appendix we have shown that this must be understood as an effect of $\cO(\Delta^2)$. It follows that the shift should be neglected if working at a leading or next-to-leading order:
\begin{equation}
\Kdf(I_\PV^s) = \Kiso(I_\PV^s) \longrightarrow \Kdf(I'^s_\PV) = \Kiso(I'^s_\PV) + \cO(\Delta^2).
\end{equation}

More generally, it follows that effects of $I_\PV^s$ may be absorbed in a redefinition of $\Kdf$ up to systematic errors at $\cO(\Delta^{k+1})$, where $k$ is the order at which we are truncating the expansion. As stressed in the main text, the estimation of the truncation error by the residual $I_\PV^s$ dependence is analogous to using scheme-dependence as an estimate of the truncation error in perturbation theory.

\section{NREFT prediction for the particle-dimer scattering length}
\label{app:NREFT}

In this appendix we explain how we obtain the 
nonrelativistic effective field theory (NREFT) prediction shown in Fig.~\ref{fig:b0}.

{
At lowest order in NREFT, the dimer-particle scattering amplitude is determined by an integral equation,
and is given in terms of the two-particle scattering length, $a_0$, 
and the three-body coupling, $H_0(\Lambda)$. 
Here $\Lambda$ is a hard cutoff introduced as an ultraviolet regularization.
The integral equation is given, for example, in Eq.~(6) of Ref.~\cite{Bedaque:1998km}.
We are interested specifically in the dimer-particle scattering length, $b_0$,
which is proportional to the scattering amplitude.
Thus we use the version of the integral equation given in Eq.~(12) of Ref.~\cite{Bedaque:1998km},
which is written for a quantity $a(k,p)$ that satisfies $a(0,0)=-b_0$.
We further rewrite this equation in terms of $b(p)= -a(0,p)$, and make 
variables dimensionless using $a_0$\footnote{In the following equations, $q $,  $m$, and $H_0$ actually denote $ q a_0$, $m a_0$ and $H_0 a_0^2$ respectively}, leading to 

\begin{align}
 \frac{b(p)}{a_0} & = -K(p,0) + \frac{2}{\pi} \int_{0}^{\Lambda } dq K(p,q) \frac{b(q)}{a_0}\,,
\label{eq:inteq}\\ 
K(p,k) &= \frac{4}{3}\left(1 + \sqrt{\frac{3p^2}{4} + 1}\right)\left[ \frac{1}{p q}\log \left( \frac{q^2+qp + p^2 + 1}{q^2-qp + p^2 + 1} \right) + \frac{2H_0}{\Lambda^2} \right]\,,
\end{align}

with the desired scattering length given by $b_0=b(0)$.
Here we have used the fact that, at the particle-dimer threshold, and in the NR limit,
\begin{equation}
-m E_{\rm NR} \equiv m (3m - E) = m \left(2m- 2m \sqrt{1-1/m^2}\right)
\approx 1\,.
\end{equation}
Given a choice of $H_0$ and $\Lambda$, 
Eq.~(\ref{eq:inteq})  can be solved by discretizing the momentum,
\begin{equation}
\int dq \to \Delta q \sum\,,
\end{equation}
with $\Delta q = \Lambda   / N_{\text{steps}}$,
and  solving the resulting matrix equation, 
\begin{equation}
b_{p} = - a_0(1 - \frac{2}{\pi}\Delta q K)_{p,k}^{-1} K_{k,0}\,.
\end{equation}
The result for $b_0$ converges sufficiently for $N_{\text{steps}}\approx 10^4$.

In order to complete the prediction we need to know the appropriate value of $H_0(\Lambda)$ to use.
This issue was addressed in Ref.~\cite{\HSrev}, where the relationship between the relativistic
quantization condition used here and the NREFT version of Ref.~\cite{\Akakib} was discussed.
In particular, Eq.~(94) of Ref.~\cite{\HSrev} shows that $H_0(\Lambda)$ vanishes if $\Kiso=0$
(as is the case here) for a choice of $\Lambda$ that is of $\cO(m)$ but is not precisely specified.
This uncertainty in $\Lambda$ is due to the difference between the smooth cutoff needed in
the relativistic quantization condition and the hard cutoff used in the NREFT approach.
The upshot is that, for one choice of $a_0$,
we need to tune the value of $\Lambda$ at which $H_0$ vanishes so that
the NREFT result for $b_0$ matches that obtained
from the relativistic quantization condition.\footnote{%
This is equivalent to fixing $\Lambda=m$ and then tuning $H_0$.}
Restoring factors of $a_0$, we do this tuning for the largest value of $m a_0$ in Fig.~\ref{fig:b0}, 
which is the most nonrelativistic case, finding $\Lambda=0.75 m$.
The results for $b_0$ at all other values of $m a_0$ are then predictions.

We expect the NREFT prediction to work well for $m a_0\gg 1$,
and this is what we find, as shown in Fig.~\ref{fig:b0}.
Indeed, this agreement provides an important cross check of the quantization condition itself
and of our numerical implementation.
The failure of the NREFT prediction that we observe for $m a_0\lesssim1$ is also expected,
and shows the importance of including relativistic effects.
}

\bibliographystyle{JHEP}      

\bibliography{ref.bib}

\end{document}